\documentclass[
    twocolumn,
    letterpaper,
    amsmath,amssymb,
    aps,prb,
    superscriptaddress, 
    floatfix
]{revtex4-2}

\usepackage{bm}        
\usepackage{slashed}   
\usepackage{braket}    
\usepackage{dcolumn}           
\usepackage{graphicx}
\graphicspath{{images/}}
\usepackage{CJKutf8}
\usepackage[percent]{overpic}
\usepackage{hyperref}
\hypersetup{
    colorlinks=true,
    linkcolor=blue,
    citecolor=blue,
    urlcolor=blue
}

\usepackage{cleveref}

\crefname{equation}{Eq.}{Eqs.}
\creflabelformat{equation}{#2(#1)#3}
\Crefname{equation}{Eq.}{Eqs.}
\crefname{figure}{Fig.}{Figs.}
\Crefname{figure}{Fig.}{Figs.}
\crefname{section}{Sec.}{Secs.}
\Crefname{section}{Sec.}{Secs.}
\crefname{appendix}{App.}{Apps.}
\Crefname{appendix}{App.}{Apps.}

\DeclareMathOperator{\Tr}{Tr}
\DeclareMathOperator{\tr}{tr}
\newcommand{\dd}{\text{d}} 
\def\I{\mathbb{I}}
\def\Z{\mathbb{Z}}

\def\L{\mathcal{L}}

\def\D{\mathcal{D}}

\def\O{\mathcal{O}}
\def\mw{\mathcal{W}}
\def\se{\mathcal{S}_{s}}
\def\hg{\hat{\mathcal{G}}}

\def\hx{\hat{\mathcal{X}}}
\def\hs{\hat{\Sigma}}
\def\hq{\hat{Q}}
\def\hv{\hat{V}}
\def\hw{\hat{W}}
\def\hm{\hat{M}}

\def\x{\boldsymbol{x}}

\def\q{\boldsymbol{q}}
\def\i{\text{i}}

\def\SO{\text{SO}}

\def\tr{\text{tr}}
\def\Tr{\text{Tr}}
\def\U{\text{U}}
\def\USp{\text{USp}}
\def\eff{\text{eff}}
\def\reg{(\text{reg})}
\def\kin{\text{kin}}

\def\BCS{\text{BCS}}

\def\diag{\text{diag}}

\def\nn{\nonumber\\&}
\def\n{\nonumber\\}
\def\en{\nonumber\\=&}
\def\c{c^\dagger}
\def\p{\psi}
\def\bp{\bar{\psi}}
\def\P{\Psi}
\def\BP{\bar{\Psi}}
\def\upa{\uparrow}
\def\doa{\downarrow}

\begin{document}

\title{Super-Logarithmic Entanglement Scaling in a Monitored Superconducting Chain}

\author{Rui-Jing Guo \begin{CJK}{UTF8}{gbsn}(郭睿婧)\end{CJK}}
\affiliation{Center for Neutron Science and Technology, Guangdong Provincial Key Laboratory of Magnetoelectric Physics and Devices, School of Physics, Sun Yat-sen University, Guangzhou 510275, China}

\author{Zhi-Yuan Wei \begin{CJK}{UTF8}{gbsn}(魏志远)\end{CJK}}
\thanks{Current Address: University of Maryland, College Park}
\affiliation{
Max-Planck-Institut f{\"{u}}r Quantenoptik, Hans-Kopfermann-Str. 1, 85748 Garching, Germany
}

\begin{abstract}
We study the entanglement dynamics of a one-dimensional spinful $s$-wave superconductor subjected to local continuous measurements. In the rare-measurement regime, we derive a replicated Keldysh non-linear sigma model (NLSM) to describe the steady-state entanglement. Within this field-theoretic framework, the interplay between measurement dephasing and superconducting pairing constrains the low-energy, long-wavelength fluctuations to an $\SO(R)$ target manifold. A one-loop renormalization-group analysis shows that the theory develops a weak-anti-localization flow, stabilizing a critical phase with super-logarithmic scaling of the steady-state entanglement. Our field-theoretic results explain the numerical evidence reported in the companion Letter~\cite{guo2026} and demonstrate that such a critical phase can emerge in a one-dimensional topologically trivial superconductor without relying on topological protection.
\end{abstract}

\maketitle

\tableofcontents
\section{Introduction}\label{sec:intro}
The interplay between unitary evolution and projective measurements in many-body quantum systems has become an active area of research for exploring non-equilibrium phases of matter~\cite{LYD_2018_PRB,Nielsen2010,Fisher_2023_random}. While unitary dynamics typically scrambles quantum information leading to volume-law entanglement growth, projective measurements induce wavefunction collapse, favoring a disentangled area-law phase. The competition between these opposing tendencies can drive a measurement-induced phase transition (MIPT), a phenomenon originally identified in quantum circuits~\cite{Jian_PhysRevB.106.134206,jian_measurementinduced_2023,Choi_2020,Agrawal_2022_PRX,Bao_2020_mar,Jian_2020_PRB,LYD_2019,Gullans_2020_PRX,Gullans_2020_PRL,Iaconis_2020,lavasani_2021,Sang_PRR,Adam_circuit_2021,Block_int_circuit,wei_2026,tikhanovskaya2023universality,lavasani_2021_Nature,Entanglement_PRX_2023}, free~\cite{fava_adam_U1,Coppola_2022_growth,Adam_PhysRevX.9.031009,Schomerus_PRB_2019,travaglino_quench_2025,swann2023spacetimepictureentanglementgeneration,chan_prb_2019,Cao_2019_SP,Ladewig_prb_2022,Carollo_2022_prb_ALBA,XheK_PRB_2023jul,Marco_PRB_2022jul,Marco_PhysRevB.105.L241114,Van_prl_MAR_2021,Poboiko_2D_PhysRevLett.132.110403,poboiko_PhysRevX.13.041046,Fava_PhysRevX.13.041045, Zhu_2024,Oshima_2025,cx_2020_emergent, Adam_major_defect,yin2026mipt,niederegger_2026_mipt,poboiko_2026_qmipt,Graham_2023_SCIPOST} and interacting~\cite{Poboiko_int,Goto_2020,Fuji_2020,Doggen_2022_PRR,Doggen_2023,foster2025interMIPT} fermionic systems, spin systems~\cite{LANG_SEP2020_APS,PRB_2021_jun_mipt,ising_2023_scipost_15,weinstein_zack_2023_jun_mipt, gal2025_Ising,Sara2023monitoredIsing,Marco2024MEEIsing}, bosonic systems~\cite{Tang_2020,pokharel2025,yang2026decoding}, disordered systems~\cite{Szyniszewski_PRB_2023_OCT}, and Sachdev-Ye-Kitaev-type models~\cite{JIAN_prl_127_2021,Altland_PhysRevResearch.4.L022066}. Furthermore, MIPTs have been experimentally realized in trapped ions~\cite{noel_measurement-induced_2022} and superconducting processors~\cite{Koh_2023,wu2025, hoke_measurement-induced_2023,Mart_n_V_zquez_2024}.

For monitored fermionic dynamics, an effective field theory based on the Keldysh closed-time contour formalism combined with the replica technique~\cite{poboiko_PhysRevX.13.041046, foster2025interMIPT} has been established. This approach treats the randomness inherent in the measurement outcomes as a source of spatiotemporal disorder, allowing for a rigorous derivation of NLSMs that describe MIPTs. Within this formalism, the MIPT is identified as a replica limit of an Anderson-like localization transition driven by spatiotemporal disorder, where the universality class is dictated by the symmetry of the underlying effective action, analogous to the Altland-Zirnbauer (AZ) classification of static disordered fermions \cite{Altland_1997,Ryu_2010, Felix_2025_keldysh}.

Recent field-theoretic investigations reveal a diverse phase diagram governed by dimensionality and symmetry. For one-dimensional free fermions with $\U(1)$ charge conservation (Class A), measurements drive the system to an area-law phase without an intervening critical point~\cite{poboiko_PhysRevX.13.041046}. Extending the study of monitored free fermions to higher dimensions ($d>1$) shows diverse critical behaviors~\cite{Poboiko_2D_PhysRevLett.132.110403}. Interactions alter this scaling, stabilizing volume-law phases through information-charge separation mechanisms~\cite{Poboiko_int}. In superconducting systems, investigations of topological Majorana $p$-wave chains (Class DIII) show that projective measurements drive the dynamics into a critical phase. An $\SO(R)$ NLSM governs this regime, producing a super-logarithmic entanglement scaling $\se \sim \ln^2 L$~\cite{Fava_PhysRevX.13.041045, foster2025interMIPT,yang2026decoding}. Beyond one-dimensional topological chains, steady-state evolution and measurements have also been demonstrated to stabilize emergent superconducting order and spin liquids in replicated spaces~\cite{SK_2025_PRL}. In contrast, the measurement-induced dynamics of a topologically trivial $s$-wave BCS chain, where the Hamiltonian and measurement action are formulated within a canonical Class CI framework, remain less explored. Our analysis shows that the topologically trivial $s$-wave system shares the same perturbative $\SO(R)$ soft-mode target as the topological $p$-wave system, but with a different dynamical mechanism and a vanishing Wess-Zumino-Witten (WZW) term. Both systems exhibit the $\se \sim \ln^2 L$ entanglement entropy scaling governed by the corresponding renormalization group flow.

In this work, we investigate a one-dimensional monitored $s$-wave BCS system~\cite{guo2026}, which is a tight-binding model including both nearest-neighbor hopping and on-site Cooper pairing, subjected to spin-resolved density measurements. This work characterizes the critical phase that emerges in the rare-measurement limit. In this model, the pairing amplitude $\Delta$ breaks the $\U(1)$ charge conservation. Unlike the topological $p$-wave case, the on-site $s$-wave pairing does not protect edge modes but instead imposes a rigid local constraint on the fermionic correlations. This places the system in a symmetry class distinct from the standard unitary and orthogonal ensembles, necessitating a careful re-examination of the symmetry-breaking pattern in replica space relative to previous studies.

The non-equilibrium steady state is shaped by the competition among three microscopic energy scales: coherent hopping $J$ generates long-range entanglement via propagating excitations~\cite{bertini_entanglement_2018}, on-site pairing $\Delta$ gaps out single-particle modes into local singlets, and local measurements $\gamma$ induce wavefunction dephasing. While measurements compete with hopping to favor quantum Zeno localization, they also suppress local pairing coherence. In the ensemble-averaged dynamics, this measurement backaction modifies the symmetry constraints on the collective fluctuation manifold, retaining soft modes that differ from those selected by pairing alone and stabilizing a critical phase rather than an area-law state.

The monitored $s$-wave BCS chain belongs to Symmetry Class CI, whose parent symplectic saddle manifold $\USp(2R)$ is projected by the measurement and pairing constraints onto an $\SO(R)$ soft sector. Renormalization group analysis in the replica limit $R \to 1$ shows a negative beta function, driving a weak anti-localization flow. This flow stabilizes a critical phase with steady-state entanglement entropy scaling as $\se \sim \ln^2 L$. No sharp phase transition occurs within the rare-measurement regime; instead, joint analytical and numerical evidence supports a critical phase throughout this regime. The system is expected to cross over toward area-law behavior when the measurement rate $\gamma$ becomes comparable to $J$ or $\Delta$, consistent with the numerical behavior observed in the companion work~\cite{guo2026}.

While pairing reduces the global $\U(1)$ charge conservation to $\Z_2$, spin-resolved density measurements preserve the total spin component $S_z$. Formulated in the canonical particle-hole basis $f = (c_\uparrow, c_\downarrow^\dagger)^{\rm T}$, the pairing interaction becomes an inter-component hybridization that preserves $S_z$. In this basis, the quadratic action respects particle-hole symmetry $\widetilde{\mathcal C} = \mu_y^{\rm PH}\mathcal K$ ($\widetilde{\mathcal C}^2 = -1$) and time-reversal symmetry $\widetilde{\mathcal T} = \mathcal K$ ($\widetilde{\mathcal T}^2 = +1$), placing the theory in symmetry class CI with a parent saddle manifold $\mathcal M_{\rm parent} \simeq \USp(2R)$. The subsequent projection from $\USp(2R)$ to the effective $\SO(R)$ target manifold is governed by the complementary mass constraints from measurement action and superconducting pairing.

The paper is organized as follows. In \cref{sec:Setup}, we introduce the Hamiltonian of the monitored BCS model and outline the replica trick formalism for computing the entanglement entropy. \cref{sec:keldysh_formalism} constructs the effective field theory through the Keldysh action and the Hubbard-Stratonovich (HS) transformation that respects the underlying symmetry constraints. In \cref{sec:Symmetry_Group}, we determine the symmetry group of the system and solve the saddle-point equations within the self-consistent Born approximation (SCBA), which identify the parent manifold of the theory. Next, \cref{sec:Fluctuation} presents the derivation of the NLSM, including the identification of the $\SO(R)$ manifold for massless modes and the analysis of the Wess--Zumino--Witten (WZW) term. Finally, in \cref{sec:RG}, we perform a renormalization group (RG) analysis to derive the $\beta$ function. This analysis shows that the system flows towards a weak-coupling fixed point, resulting in a critical phase characterized by a super-logarithmic ($\ln^2 L$) scaling of the entanglement entropy.

\section{Microscopic Model and Replica Formalism}\label{sec:Setup}
\subsection{Hamiltonian and Measurement Protocol}\label{sec:Hamiltonian}
We study a one-dimensional spin-1/2 $s$-wave BCS lattice consisting of $L$ sites. The system is governed by the Hamiltonian, which includes nearest-neighbor hopping of strength $J$ and on-site fermion pairing term with the pairing amplitude $\Delta$:
\begin{align} \label{eq:hamil}
    H_\BCS &= -J\sum_{\sigma\in\{\upa,\doa\}}\sum_{j=1}^{L}\left(\c_{j,\sigma}c_{j+1,\sigma}+\text{H.c.}\right) 
    \nn \quad -\Delta\sum_{j=1}^L \left( \c_{j,\upa}\c_{j,\doa}+\text{H.c.}\right),
\end{align}
Here, the operators $c^\dagger_{j,\sigma}$ and $c_{j,\sigma}$ denote the creation and annihilation of a fermion with spin $\sigma \in \{\upa, \doa\}$ at site $j\in [1,L]$, respectively. We set the chemical potential to zero, corresponding to half filling, which simplifies the particle-hole relations and enables chiral-like cancellations~\cite{Ryu_2010}. We take the pairing amplitude $\Delta$ to be real, as a global $\U(1)$ gauge transformation can remove any complex phase. The pairing term reduces the global $\U(1)$ charge conservation to a discrete $\Z_2$ fermion-parity symmetry. We also impose periodic boundary conditions (PBC), defined by the relation $c_{L+1,\sigma}\equiv c_{1,\sigma}$. 
We use specific initial configurations, such as the N\'eel product state and the vacuum state, for numerical calculations in~\cite{guo2026}, whereas this NLSM framework establishes a universal description independent of the initial conditions for the long-time, long-wavelength steady-state scaling considered here.

The system also undergoes random local projective measurements of the spin-resolved occupation numbers $\hat{n}_{j,\sigma} = \c_{j,\sigma} c_{j,\sigma}$. A measurement event at site $j$ projects the local quantum state into one of the four basis states within the local Fock space: $\ket{0}_j,\ket{\upa}_j,\ket{\doa}_j,\ket{\upa\doa}_j$. The projection operator $\hat{P}_{\mathbf{n}_j}(j)$ corresponding to the outcome tuple $\mathbf{n}_j = (n_{j,\upa}, n_{j,\doa})$, where the measured particle number $n_{j,\sigma} \in \{0,1\}$, is given by:
\begin{equation}\label{eq:projector_def}
    \hat{P}_{\mathbf{n}_j}(j) = \prod_{\sigma\in\{\upa,\doa\}} \left[ n_{j,\sigma} \hat{n}_{j,\sigma} + (1-n_{j,\sigma})(1-\hat{n}_{j,\sigma}) \right].
\end{equation}
These local projectors satisfy the standard completeness relations and orthogonality relations, as 
\begin{equation} \label{eq:P_nj}
	\sum_{\mathbf{n}} \hat{P}_{\mathbf{n}}(j) = \I, \quad \hat{P}_{\mathbf{n}}(j) \hat{P}_{\mathbf{n}'}(j) = \delta_{\mathbf{n}, \mathbf{n}'} \hat{P}_{\mathbf{n}}(j).
\end{equation}
We model these measurement events as independent Poisson processes occurring at each lattice site. Within an infinitesimal time interval $\delta t\ll1$, a measurement takes place with probability $p = \gamma \delta t$, where $\gamma$ defines the measurement rate. Therefore, a chain of length $L$ undergoing a total evolution time $T$ yields an expected total number of measurement events $\overline{M} = \gamma L T$.

In the rare-measurement regime $\gamma \ll \Delta, J$, the long-wavelength dynamics is governed by the interplay of measurement dephasing and pairing gap constraints. Numerical evidence of measurement-induced entanglement enhancement was reported in the companion Letter~\cite{guo2026}; here we establish the continuum field-theoretic description of the steady state.

At long wavelengths, the cumulative backaction of the discrete projections is coarse-grained into a continuous spatiotemporal disorder potential that renormalizes the effective stiffness and selects the target manifold of the NLSM. While the clean bulk quasiparticle spectrum possesses a pairing gap $\Delta$ that eliminates single-particle excitations at zero energy, the low-energy excitations governing the NLSM are Goldstone modes in Keldysh-replica space rather than fermionic quasiparticles. In the ensemble-averaged theory, stochastic measurement dephasing broadens the low-energy spectral function, generating a finite subgap spectral weight $\nu_{\rm sub} \sim \nu_F \gamma / \Delta$. This subgap density of states provides a non-zero stiffness for the collective orientational fluctuations, enabling these massless replica modes to mediate super-logarithmic entanglement scaling.

We formulate the statistical properties of the Poissonian measurement protocol in the following subsection.

\subsection{Statistical Ensemble and Replica Trick}\label{sec:replica}
We consider the evolution of the system starting from an initial density matrix $\hat{\rho}_{\text{in}}$. The dynamics consist of alternating intervals of coherent unitary propagation, $\hat{U}_0(t_{m+1}, t_m) = e^{-\i H_\BCS (t_{m+1}-t_m)}$, and instantaneous projective measurements at times $\{t_m\}$. Define the evolution of the unnormalized density matrix $\hat{D}(t)$, which originates from $\hat{D}(0) = \hat{\rho}_{\text{in}}$, as: 
\begin{align} \label{eq:evolution_D}
    \hat{D}(t_{m+1}-0) &= \hat{U}_0(t_{m+1}, t_m) \hat{D}(t_m+0) \hat{U}_0^\dagger(t_{m+1}, t_{m}), \n
    \hat{D}(t_m + 0) &= \hat{P}_{\mathbf{n}_m}(j_m) \hat{D}(t_m - 0) \hat{P}_{\mathbf{n}_m}(j_m),
\end{align}
where $\hat{P}_{\mathbf{n}_m}(j_m)$ is the local projector defined in \cref{eq:projector_def}, corresponding to a specific measurement outcome $\mathbf{n}_m$ at site $j_m$.

The relation $\hat{\rho}(t) = \hat{D}(t)/\Tr\hat{D}(t)$ defines the normalized density matrix for a single realization. The trace $\Tr\hat{D}(T)$ gives the Born probability weight for the measurement trajectory. We define the kinematic Poisson average $\overline{\O}_{\text{kin}}$ for a trajectory-dependent observable by summing over all measurement outcomes $\{\mathbf{n}_m\}$ and integrating over the Poissonian distribution of measurement times $\{t_m\}$:
\begin{align}\label{eq:overline_def}
    \overline{\mathcal{O}}_{\text{kin}} \equiv
    \sum_{M=0}^{\infty} P(M)
    \left[ \sum_{\{\mathbf{n}_m\}} \prod_{m=1}^{M} \sum_{j_m=1}^{L} \int_{0}^{T} \frac{\dd t_m}{TL} \right] \mathcal{O}.
\end{align}
Here, $P(M) = e^{-\overline{M}}(\overline{M})^M / M!$ weights the contribution of $M$ measurement events, and the measure $1/(TL)$ accounts for the uniform spatial and temporal distribution of each independent event. The trajectory average (ensemble average) $\mathbb{E}[\mathcal{O}]$ is defined by weighting over the trajectory probability:
  \begin{equation}
      \mathbb{E}[\mathcal{O}] = \sum_{\rm traj} P_{\rm kin}^{[{\rm traj}]}\, p_{\rm Born}^{[{\rm traj}]}\, \mathcal{O}^{[{\rm traj}]},
  \end{equation}
  where the Born probability weight is $p_{\rm Born}^{[{\rm traj}]} = \mathrm{Tr}\,\hat{D}^{[{\rm traj}]}$, and the kinematic trajectory probability is $P_{\rm kin}^{[{\rm traj}]} = P(M) / (TL)^M$.

We compute the replica moments for integer $R \ge N$, analytically continue in $R$ and $N$, evaluate the normalization replica limit at $R \to 1$ in the analytically continued expression, and then take $N \to 1$ to recover the von Neumann entanglement entropy. The ensemble-averaged replicated density matrix is given by:
\begin{align}
    \hat{\rho}_N &= \mathbb{E}\left[ \hat{\rho}^{\otimes N} \right] = \sum_{\text{traj}} P_{\text{kin}}^{[{\rm traj}]} p_{\text{Born}}^{[{\rm traj}]} \frac{\hat{D}^{\otimes N}}{(\Tr \hat{D})^N}
    \nn = \sum_{\text{traj}} P_{\text{kin}}^{[{\rm traj}]} (\Tr \hat{D}) \frac{\hat{D}^{\otimes N}}{(\Tr \hat{D})^N}
    \nn= \sum_{\text{traj}} P_{\text{kin}}^{[{\rm traj}]} \frac{\hat{D}^{\otimes N}}{(\Tr \hat{D})^{N-1}}
     = \overline{ \frac{\hat{D}^{\otimes N}}{(\Tr\hat{D})^{N-1}} }_{\text{kin}}.
\end{align}
The non-linear trajectory dependence of the normalization factor $(\Tr\hat{D})^{-N+1}$ makes the direct calculation difficult. Thus we use the replica trick to compute $\hat{\rho}_N$~\cite{edwards_theory_1975,ma2023critical,wegner_mobility_1979,poboiko_PhysRevX.13.041046}. By introducing an integer number of replicas $R \ge N$ and performing a partial trace over the extra $R-N$ copies, we absorb the denominator into a unified statistical weight. Under the replica construction with $R$ replicas, the partial trace over replicas $N+1$ to $R$ leaves $N$ copies in the numerator, while the overall trace in the $R \to 1$ limit naturally absorbs the Born probability factor $(\Tr\hat{D})^{-N+1}$ under the kinematic Poisson average. The analytic continuation to $R \to 1$ recovers the normalized average:
\begin{align}\label{eq:replica_def}
    \hat{\rho}_N = \lim_{R\to 1}\Tr_{N+1,\dots,R}
    \bigg[\,\overline{\bigotimes_{r=1}^{R}\hat{D}_r(T)}_{\text{kin}}\,\bigg].
\end{align}
We describe the simultaneous evolution of these $R$ replicas within the Keldysh path-integral formalism, applying the contour-ordering operator $\mathcal{T}_C$:
\begin{align}\label{eq:time_ordering}
    \bigotimes_{r=1}^R \hat{D}_r(T) &= \mathcal{T}_C \Bigg[ e^{-\i\int_C \dd t\, H_\BCS} \prod_{m=1}^M \bigotimes_{r=1}^R 
    \nn\quad \times \left( \hat{P}_{\mathbf{n}_m, r}^+(j_m, t_m) \hat{P}_{\mathbf{n}_m, r}^-(j_m, t_m) \right) \Bigg] \hat{\rho}_{\text{in}}^{\otimes R}.
\end{align}
Since a given trajectory realizes a specific sequence of measurement outcomes $\{\mathbf{n}_m\}$, the corresponding projection operators at each event act identically on the forward ($+$) and backward ($-$) branches of the Keldysh contour. This shared outcome constrains the field configurations on the contour. Subsequently, performing the ensemble average over all possible outcomes and Poissonian events, as defined in \cref{eq:overline_def}, generates an effective coupling among the $R$ distinct replicas.

To compute the von Neumann entanglement entropy for a spatial subsystem $A$ of length $L_A$, one evaluates the moments of the reduced density matrix, $\Tr_A(\hat{\rho}_A^N)$. Within the replica framework~\cite{Pasquale_Calabrese_2004,poboiko_PhysRevX.13.041046,Calabrese_2009,Jian_2020_PRB}, the moment $\Tr_A(\hat{\rho}_A^N)$ is obtained by inserting the subsystem replica permutation operator $\hat{\mathcal{T}}_{A}^{(N)}$ that shifts replica indices cyclically inside region $A$, generating twist defect configurations at the boundary points [cf.~\cref{sec:EE}]. The von Neumann entropy is extracted through analytic continuation in the replica index $N$:
\begin{align} \label{eq:entropy_def}
\overline{\se}(T) = -\lim_{N\to 1} \frac{\partial}{\partial N} \ln \Tr \left( \hat{\rho}_N \hat{\mathcal{T}}_{A}^{(N)} \right).
\end{align}
This formulation exposes two distinct replica limits governing the underlying physics: the limit $R \to 1$ implies the conservation of probability and determines the symmetry of the NLSM manifold, while $N \to 1$ governs the scaling behavior of the entanglement entropy. We define the sequence of analytic continuations for the replica calculation: all quantities are evaluated for integer replica numbers satisfying $R \ge N \ge 1$. The results are then continued to real values of $R$ and $N$. We take the normalization limit $R \to 1$ first to preserve probability conservation, followed by the limit $N \to 1$ to obtain the von Neumann entropy.

The independent Poissonian distribution of measurement events introduces a separation of scales. At short length and time scales, the unitary evolution of fermions between isolated measurements determines the system dynamics. In the long-wavelength limit, averaging the non-linear backaction of these discrete projections over all trajectories within the Keldysh formalism maps the measurement protocol to a continuous spatiotemporal disorder potential~\cite{Felix_2025_keldysh,Barratt_2022,poboiko_PhysRevX.13.041046}.

\section{Fermionic Field Theory and Basis}
\label{sec:keldysh_formalism}

This section constructs the replicated Keldysh action used in the saddle-point analysis. We first rewrite the BCS Hamiltonian in a canonical particle-hole basis and then average the shared measurement outcomes on the Keldysh contour. We finally introduce slow matrix fields to obtain the collective action studied in
\cref{sec:Symmetry_Group}.

To exploit the conservation of $S_z$, we employ the canonical particle-hole transformation
\begin{align}\label{eq:f_basis}
    f_j\equiv
    \begin{pmatrix}
        f_{j,1}\\
        f_{j,2}
    \end{pmatrix}
    =
    \begin{pmatrix}
        c_{j,\uparrow}\\
        c_{j,\downarrow}^\dagger
    \end{pmatrix},
    \quad
    f_k\equiv
    \begin{pmatrix}
        c_{k,\uparrow}\\
        c_{-k,\downarrow}^\dagger
    \end{pmatrix},
\end{align}
which maps the two local electron modes onto a canonical doublet without introducing redundant Nambu copies. Under this transformation, the pairing Hamiltonian becomes an on-site inter-component hybridization $-\Delta \sum_j f_j^\dagger \mu_x^{\rm PH} f_j$ that preserves $S_z$ [cf.~\cref{ap:Keldysh}].

We use the labels ${\rm K}$, ${\rm PH}$, and ${\rm R}$ for the Keldysh, particle-hole, and replica spaces. The matrices $\tau_i^{\rm K}$ and $\mu_i^{\rm PH}$ act in the Keldysh and particle-hole spaces, respectively, where $i\in\{0,x,y,z\}$. The matrices $\tau_0^{\rm K}$ and $\mu_0^{\rm PH}$ denote the corresponding identity
matrices. The spin labels are incorporated into the particle-hole doublet, so the replicated field has no separate spin-space factor. The replica index is
$r\in\{1,\ldots,R\}$, and $\I_R$ denotes the identity in
replica space.
The charge-conjugation structure defines the antisymmetric form
\begin{align}\label{eq:metric}
    \hx= \tau_0^{\rm K}\otimes \mu_y^{\rm PH}\otimes\I_R,
    \quad
    \hx^{\rm T} =-\hx.
\end{align}
This form enters the symmetry constraints on the replicated matrix fields in \cref{sec:Symmetry_Group}.

After applying the LO rotation defined in \cref{ap:Keldysh}, the operator doublet is represented by the independent complex Grassmann fields
\begin{subequations}\label{eq:PH_spinor_explicit}
\begin{align}
    \P_{b,r}
    &=
    \begin{pmatrix}
        f_{1,b,r}\\
        f_{2,b,r}
    \end{pmatrix}
    =
    \begin{pmatrix}
        \p_{\uparrow,b,r}\\
        \bp_{\downarrow,b,r}
    \end{pmatrix},
    \\
    \BP_{b,r}
    &=
    \begin{pmatrix}
        \bar f_{1,b,r}&
        \bar f_{2,b,r}
    \end{pmatrix}
    =
    \begin{pmatrix}
        \bp_{\uparrow,b,r}&
        \p_{\downarrow,b,r}
    \end{pmatrix}.
\end{align}
\end{subequations}
Here, $b\in\{1,2\}$ labels the two LO components, $\alpha\in\{1,2\}$ labels the particle-hole components, and $r\in\{1,\ldots,R\}$ labels the replicas. We assemble these indices into the composite field
\begin{align}\label{eq:composite_field_main}
    \P_a=(\P_{b,r})_\alpha,
    \quad
    \BP_a=(\BP_{b,r})_\alpha,
\end{align}
where $a=(b,\alpha,r)$. Thus, $\P$ and $\BP$ each contain $4R$ Grassmann components and are organized in Keldysh $\otimes$ particle-hole $\otimes$ replica space. The barred field $\BP$ is independent of $\P$ and is not its Hermitian conjugate. The replicated Keldysh action for the monitored BCS system over the evolution interval $[t_i,t_f]$ is
\begin{align}\label{eq:S_psi}
\i S[\BP,\P] = \i S_0[\BP,\P] + \gamma\int_{t_i}^{t_f}\dd t\int\dd x\, \bigl( \i\L_M[\BP,\P] \bigr).
\end{align}
Here, $\x=(t,x)$, $t_i$ is the initial time, $t_f$ is the observation time at which the observable or replica twist is inserted, and $\gamma$ is the measurement rate. In the translation-invariant bulk analysis, we take $t_f-t_i\to\infty$ and omit contributions localized near the temporal endpoints. Below, $\int\dd^2\x$ in a measurement term denotes the same bounded spacetime integral.

The quadratic action in the full replicated space is
\begin{align}\label{eq:L0_clean}
    \i S_0=\i\int\frac{\dd k}{2\pi}\frac{\dd\omega}{2\pi}\,\BP(k,\omega)G_0^{-1}(k,\omega)\P(k,\omega)
\end{align}
where the single-replica Hamiltonian $h_f(k)$ in the basis is
\begin{align}\label{eq:Hamiltonian_k}
    h_f(k)= \xi_k\mu_z^{\rm PH}-\Delta\mu_x^{\rm PH},
    \quad
    \xi_k=-2J\cos k.
\end{align}
The replicated inverse Green's function is
\begin{align}
    G_0^{-1}(k,\omega)\equiv \tau_0^{\rm K} \otimes\left[ \omega\mu_0^{\rm PH}-h_f(k)\right] \otimes\I_R+\i\delta\Lambda_0(\omega),
\end{align}
The infinitesimal term $\i\delta\Lambda_0(\omega)$ with $\delta\to0^+$ records the initial-state boundary condition. The distribution function $F_0(\omega)=1-2f_0(\omega)$ enters the off-diagonal Keldysh component of $\Lambda_0(\omega)$, where $f_0(\omega)$ is the initial single-particle mode occupation. For the translation-invariant long-time bulk calculation, where temporal-boundary transients are omitted, we choose the reference causal regulator $F_0(\omega)=0$, corresponding to $f_0(\omega)=1/2$ mode by mode. With this choice, the causal matrix simplifies to
\begin{align}\label{eq:lambda_0}
    \Lambda_0= \tau_z^{\rm K}\otimes\mu_0^{\rm PH} \otimes \I_R.
\end{align}
This choice defines the bare reference regulator used in the bulk calculation, while the long-time measurement-averaged distribution is determined independently by the saddle-point equations in \cref{sec:SCBA}. More general matrix-valued boundary distributions are discussed in \cref{ap:Keldysh}.

At a fixed spacetime measurement event, the outcome $\mathbf n=(n_\uparrow,n_\downarrow)\in\{(0,0),(0,1),(1,0),(1,1)\}$ acts on both contour branches and on all replicas. In the normal-ordered coherent-state representation, this event is described by the local Lagrangian density
\begin{align}\label{eq:Lm}
    \i\L_M[\BP,\P]=\sum_{\mathbf n}\left[ \prod_{r=1}^{R}(\hv_{\mathbf n})_{+,r}[\BP_r,\P_r]\,(\hv_{\mathbf n})_{-,r}[\BP_r,\P_r]\right]-1.
\end{align}
The sum over $\mathbf n$ accounts for the shared measurement outcome across contour branches and replicas. The constant term $-1$ arises from the Poisson no-event weight $e^{-\overline{M}}=e^{-\gamma LT}$ in the generating functional, which preserves normalization and causality. Here $\i\L_M[\BP,\P]$ represents a real dissipative kernel in the exponent, and the measurement contribution is $\i S_M[\BP,\P]=\int\dd^2\x\,(\i\L_M[\BP,\P])$. For the projector $\hat P_{\mathbf n}$ defined in \cref{eq:projector_def}, its unregularized coherent-state symbol on branch $\pm$ and replica $r$ is $(\hv_{\mathbf n})_{\pm,r}\equiv\langle\bp_{\pm,r}|\hat P_{\mathbf n}|\p_{\pm,r}\rangle/\langle\bp_{\pm,r}|\p_{\pm,r}\rangle$. These branch fields are expressed through the LO-rotated variables $\P_r$ and $\BP_r$ that enter \cref{eq:L0_clean}, as detailed in \cref{ap:Keldysh}.

Because the microscopic projectors are normal ordered, evaluating coincident-time products in the path integral requires symmetric equal-time regularisation. Matching the resulting local contact counterterms re-expresses the measurement Lagrangian density in the regularised form in \cref{eq:L_M_final_res_strict}, which yields identical perturbative correlation functions (see \cref{ap:Keldysh}).

At the operator level, the transformation $f_2=c_\downarrow^\dagger$ gives
\begin{align}
    \hat n_\uparrow+\hat n_\downarrow=1+f^\dagger\mu_z^{\rm PH}f,
    \quad
    \hat n_\uparrow-\hat n_\downarrow=-1+f^\dagger\mu_0^{\rm PH}f.
\end{align}
Here, $f$ and $f^\dagger$ denote fermionic operators, and the constant shifts originate from $c_\downarrow c_\downarrow^\dagger=1-\hat n_\downarrow$. Mapping these operators to regularised coherent-state symbols and performing the LO rotation yields the contour-symmetric measurement matrices
\begin{align}\label{eq:hm12_def}
    \hm_c=\tau_x^{\rm K}\otimes\mu_z^{\rm PH}\otimes\I_R,
    \quad
    \hm_s=\tau_x^{\rm K}\otimes\mu_0^{\rm PH}\otimes\I_R.
\end{align}
The matrices $\hm_c$ and $\hm_s$ describe local charge-density fluctuations and conserved spin-imbalance fluctuations, respectively. Summing over the four measurement outcomes symmetrizes these fluctuations and yields
\begin{align}\label{eq:L_M_final_res_strict}
    \i\L_M[\BP,\P]=\frac{2}{4^{2R}} \sum_{\beta\in\{c,s\}}\cosh\!\left(2\BP\hm_\beta\P \right)-1.
\end{align}
For each particle-hole flavor, the product of the regularized branch-pair vertices over the $R$ replicas yields a factor $4^{-R}$, producing the overall prefactor $4^{-2R}$. The regularisation steps, the outcome sum, and the channel decomposition are detailed in \cref{ap:Keldysh}.

To reformulate the fermionic interaction in terms of collective variables, we introduce the local equal-time bilinear field $\hg$ and its conjugate self-energy field $\hs$. The field $\hs$ enforces the constraint $\hg_{ab}=-\i\P_a\BP_b$ on the fermionic bilinear. In the long-wavelength matrix theory, $\hg$ is projected onto small center-of-mass frequencies and momenta. Rewriting the measurement interaction in terms of this slow field requires retaining all contraction channels that form slow fermionic bilinears.
The constraint is imposed via a Hubbard--Stratonovich transformation by integrating over the matrix fields $\hg$ and $\hs$. Here, $\Tr(\cdots)\equiv\int\dd^2\x\,\tr(\cdots)$ contains the spacetime integration, while the lower-case trace acts on Keldysh, particle-hole, and replica indices. The convergent Gaussian formulation of this identity and the corresponding integration contours are detailed in \cref{ap:GHS}.

Integrating out the Grassmann fields yields the Gaussian collective action
\begin{align}\label{eq:S0_hs_hg}
    \i S_{0}[\hg,\hs]=\Tr\ln\!\left(G_0^{-1}+\i\hs\right)-\i\Tr\!\left(\hs\hg\right).
\end{align}
The high-transfer components are integrated out, with their local contributions absorbed into the parameters of the low-energy action. From this point onward, $\hg$ and $\hs$ denote the retained slow fields. The local constraint defining $\hg$ determines one fermionic bilinear but does not select a unique pairing pattern in higher-order measurement vertices. The low-energy vertex includes every contraction channel where all transferred frequencies and momenta remain within the slow momentum window. In \cref{ap:GHS}, we define these disjoint pairing regions and show by phase-space power counting that their sum corresponds to the cycle decomposition of the symmetric group $\mathfrak{S}_m$ generated by the normalized Gaussian average with covariance $\hg$. Applying this all-channel construction to the measurement interaction in \cref{eq:L_M_final_res_strict} gives the low-energy effective density
\begin{align}\label{eq:LM_det_strict}
    \i\L_M[\hg]\simeq\sum_{\beta\in\{c,s\}}\sum_{\eta=\pm1}\det_{4R}\!\left(\frac{\I}{2}-\i\eta\hg\hm_\beta\right)-1.
\end{align}
Here, the determinant acts on the $4R$-dimensional Keldysh $\otimes$ particle-hole $\otimes$ replica space, absorbing the prefactor $4^{-2R}$ via $\det_{4R}(A/2)=4^{-2R}\det_{4R}A$. This determinant representation is an asymptotic low-energy approximation controlled by the cutoff ratio $\lambda\ll1$. As demonstrated in \cref{ap:GHS}, the $\mathcal{O}(\lambda)$ error bound is uniform with respect to the replica index $R$ and remains stable under the continuation $R\to 1$.

After coarse graining over scales large compared with the lattice spacing, the total Keldysh action takes the form
\begin{align}\label{eq:S_hs_hg}
\i S[\hg,\hs] = \i S_{0}[\hg,\hs] + \gamma\int\dd^2\x\,\bigl( \i\L_M[\hg] \bigr).
\end{align}
This action serves as the starting point for the saddle-point equations in \cref{sec:SCBA}. The stationary orientations of $\hg$ and $\hs$ determine the parent manifold analyzed in \cref{sec:parent_manifold}.

\section{Saddle Point Analysis and Parent Manifold}\label{sec:Symmetry_Group}

For the reduced Hamiltonian in \cref{eq:Hamiltonian_k}, the particle-hole and time-reversal operations can be chosen as
\begin{align}
    \widetilde{\mathcal C}=\mu_y^{\rm PH}\mathcal K,
    \quad
    \widetilde{\mathcal T}=\mathcal K.
\end{align}
They satisfy
\begin{align}
    \widetilde{\mathcal C}h_f(k)\widetilde{\mathcal C}^{-1}
    &=\mu_y^{\rm PH}h_f^*(k)\mu_y^{\rm PH} =
    -h_f(-k),\,\,
    \widetilde{\mathcal C}^{\,2}=-1,
    \label{eq:PHS_explicit}
    \\
    \widetilde{\mathcal T}h_f(k) \widetilde{\mathcal T}^{-1}
    &= h_f^*(k)=  h_f(-k),
    \,\,
    \widetilde{\mathcal T}^{\,2}= +1.
    \label{eq:TRS_explicit}
\end{align}
Their product gives the unitary chiral operation
\begin{align}
    \widetilde{\mathcal S} = \widetilde{\mathcal C}\widetilde{\mathcal T}= \mu_y^{\rm PH},
    \quad
    \left\{\widetilde{\mathcal S}, h_f(k) \right\} = 0.
\end{align}
These relations place the quadratic kernel in symmetry class CI.

The measurement matrices obey
\begin{align}
   &\widetilde{\mathcal T}\hm_\beta \widetilde{\mathcal T}^{-1}=\hm_\beta,
    \quad
    \beta\in\{c,s\},
    \nn \widetilde{\mathcal C}\hm_c\widetilde{\mathcal C}^{-1}=-\hm_c, 
    \quad\widetilde{\mathcal C} \hm_s\widetilde{\mathcal C}^{-1}=\hm_s.
\end{align}
The sign in the charge channel is absorbed by the change of variable $\eta\mapsto-\eta$ in the outcome sum in \cref{eq:LM_det_strict}. Hence, the outcome-averaged measurement action remains compatible with the class-CI anti-unitary structure.

The anti-unitary classification fixes the compact symplectic structure of the class-CI replica saddle. The independent left and right replica transformations satisfy
\begin{align}
    u_a^\dagger u_a=\I_{2R},
    \quad u_a^{\rm T}\left(\mu_y^{\rm PH}\otimes\I_R\right)u_a=\mu_y^{\rm PH}\otimes\I_R,
\end{align}
where $a\in\{{\rm L},{\rm R}\}$.
The parent symmetry group is therefore
\begin{align}\label{eq:parent_symmetry_G}
    \mathbb G_{\rm CI}=\USp(2R)_{\rm L} \times\USp(2R)_{\rm R}.
\end{align}
These factors act by left and right multiplication on the replica group field constructed in \cref{sec:parent_manifold}.

\subsection{Saddle-Point Equations}\label{sec:SCBA}

For a translation-invariant saddle, the stationary conditions of
\cref{eq:S_hs_hg} are
\begin{subequations}\label{eq:SCBA_main}
    \begin{align}
        \hg_{\mathrm{local},0}
        &=
        \int_{-\pi}^{\pi}\frac{\dd k}{2\pi}
        \int_{-\infty}^{\infty}\frac{\dd\omega}{2\pi}
        \left[  G_0^{-1}(k,\omega)+\i\hs_0\right]^{-1},
        \label{eq:saddle_dyson}
        \\
        0&=-\i\hs_0+\gamma\left.\frac{ \delta\!\left(\i\L_M[\hg]\right)}{\delta\hg }\right|_{\hg=\hg_{\mathrm{local},0}}.
        \label{eq:saddle_measurement}
    \end{align}
\end{subequations}
The first equation is the symmetric equal-time Green's function defined by \cref{eq:G_reg_def}, while the second equation retains the full measurement determinant in \cref{eq:LM_det_strict}. The coupled saddle equations are evaluated in \cref{ap:saddle_point}.

For the normalized saddle parametrization, with $\hq_0$ defined in
\cref{eq:ansatz_q0},
\begin{align}
    \hg_{\mathrm{local},0}= -\frac{\i}{2}\hq_0.
\end{align}
The matrices at the reference saddle obey
\begin{align}
    \left(\hq_0\hm_\beta\right)^2=-\I_{4R},
    \quad
    \det_{4R}\!\left(\I_{4R}-\eta\hq_0\hm_\beta\right)= 4^R,
\end{align}
where $\beta\in\{c,s\}$ and $\eta=\pm1$.
The first variation of \cref{eq:LM_det_strict} is
\begin{align}
    \left. \delta\!\left( \i\L_M\right)\right|_{\hg_{\mathrm{local},0}}=  \i4^{1-R}\tr\!\left( \hq_0\,\delta\hg\right).
\end{align}
\Cref{eq:saddle_measurement} then gives
\begin{align}
    \hs_0 = \gamma 4^{1-R}\hq_0 \xrightarrow{R\to1}\gamma\hq_0.
\end{align}
Under the unconditioned ensemble average, the continuous monitoring generates a unital channel. Because both total $S_z$ and fermion parity commute with the Hamiltonian and the measurement operators, the dynamics decompose into disconnected symmetry sectors. For an initial state in the half-filled sector with total $S_z=0$ and even parity, monitoring drives the unconditioned state toward the maximally mixed state restricted to that conserved subspace, while single conditional trajectories remain pure~\cite{poboiko_PhysRevX.13.041046,Jin_2022_PRR}. In this invariant sector, local single-particle occupations average to $f_0=1/2$, establishing the causal structure of the saddle point. In Keldysh space, the collective matrix $\hg$ contains retarded ($\hg^R$), advanced ($\hg^A$), and Keldysh ($\hg^K$) components related by $\hg^K = \hg^R \hat{F} - \hat{F} \hg^A$. The sector-restricted uniform distribution corresponds to a vanishing distribution matrix $\hat{F}=0$, which sets the off-diagonal component $\hg^K=0$.

The infinite-temperature distribution selects the reference causal orientation
\begin{align}\label{eq:ansatz_q0}
    \hq_0=\tau_z^{\rm K} \otimes \mu_0^{\rm PH}\otimes \I_R.
\end{align}
The retarded and advanced components carry opposite signs in this configuration. The self-energy amplitude and its particle-hole structure are fixed by \cref{eq:SCBA_main}.

The normalized orientational fields are parameterized as
\begin{align}
    \hs(\x)=\gamma 4^{1-R}\hq(\x),
    \quad
    \hg_{\mathrm{local}}(\x)= -\frac{\i}{2}\hq(\x),
\end{align}
and satisfy $\hq^2(\x)=\I_{4R}$. With the convention of \cref{eq:S0_hs_hg}, the fermion kernel is
\begin{align}
    G_0^{-1}+\i\hs= G_0^{-1}+\i\gamma 4^{1-R}\hq.
\end{align}
The cross term $-\i\Tr(\hs\hg)$ is independent of the orientation.
After omitting this constant term, the orientational trace-log action is
\begin{align}\label{eq:S0_hq_final}
    \i S_{0}[\hq]=\Tr\ln\!\left(G_0^{-1}+ \i\gamma 4^{1-R}\hq\right).
\end{align}
Substitution of $\hg=-\i\hq/2$ into \cref{eq:LM_det_strict} gives
\begin{align}\label{eq:SM_hq_final}
    \i S_M[\hq]=\int\dd^2\x\left[\sum_{\beta\in\{c,s\}}\sum_{\eta=\pm1}\det_{4R}\!\left(\frac{\I_{4R}-\eta\hq\hm_\beta}{2}\right)-1\right].
\end{align}
The orientational action is
\begin{align}\label{eq:S_hq_total}
\i S[\hq] = \i S_{0}[\hq] + \gamma \bigl( \i S_M[\hq] \bigr).
\end{align}
The local mass matrix is defined by the second variation of the total orientational action $\i S[\hq]$ around the reference saddle. Its eigenvectors determine the massive and soft directions within the class-CI tangent space.

\subsection{Parent Manifold}\label{sec:parent_manifold}

A fixed rotation in Keldysh space maps the causal saddle in \cref{eq:ansatz_q0} to
\begin{align}
    \overline{\hq}_0 =
    \begin{pmatrix}
        0 & \I_{2R}\\
        \I_{2R} & 0
    \end{pmatrix}.
\end{align}
For $(u_{\rm L},u_{\rm R})\in\mathbb G_{\rm CI}$, the orbit of this reference configuration is
\begin{align}
    \overline{\hq}[g]
    &=
    \begin{pmatrix}
        u_{\rm L} & 0\\
        0 & u_{\rm R}
    \end{pmatrix}
    \overline{\hq}_0
    \begin{pmatrix}
        u_{\rm L}^{-1} & 0\\
        0 & u_{\rm R}^{-1}
    \end{pmatrix}
    \nn=
    \begin{pmatrix}
        0 & g\\
        g^{-1} & 0
    \end{pmatrix},
    \quad
    g=u_{\rm L}u_{\rm R}^{-1}.
\end{align}
Closure of the unitary symplectic group gives
\begin{align}
    g^\dagger g=\I_{2R},
    \quad g^{\rm T}\left(\mu_y^{\rm PH}\otimes\I_R\right)g
    =\mu_y^{\rm PH}\otimes\I_R,
\end{align}
so that $g\in\USp(2R)$. The embedded matrix field satisfies
\begin{align}
    \overline{\hq}[g]^\dagger=\overline{\hq}[g],\quad
    \overline{\hq}[g]^2=\I_{4R}.
\end{align}
The reference saddle remains invariant when
$u_{\rm L}u_{\rm R}^{-1}=\I_{2R}$, or equivalently
$u_{\rm L}=u_{\rm R}$. Its stabilizer is therefore
\begin{align}
    \mathbb H_{\rm CI} = \USp(2R)_{\rm diag}.
\end{align}
The orbit of the saddle is the parent manifold
\begin{align}\label{eq:coset_parent_modified}
    \mathcal M_{\rm parent}
    &=\frac{\USp(2R)_{\rm L}\times\USp(2R)_{\rm R}}{\USp(2R)_{\rm diag}}
    \simeq\USp(2R).
\end{align}
Rotating $\overline{\hq}[g]$ back to the causal Keldysh basis parameterizes the orientational fluctuations on the parent manifold $\mathcal M_{\rm parent}$ by $\hw = \tau_x^{\rm K} \otimes W_1$ with $W_1 \in \mathfrak{usp}(2R)$ [cf.~\cref{ap:G_hw}]. The trace-log in \cref{eq:S0_hq_final} and the measurement determinant in \cref{eq:SM_hq_final} define the orientational action on $\mathcal M_{\rm parent}$.

\section{Fluctuation Analysis}\label{sec:Fluctuation}
\subsection{Symmetry Analysis and NLSM Manifold}\label{sec:SOR}
We parameterize the fluctuations of the matrix field $\hq(\x)$ around the saddle point $\hq_0$ through a space-time dependent unitary rotation $U(\x) =\exp[\hw(\x)/2]$ as:
\begin{align}\label{eq:exp_param}
\hq(\x) = U(\x) \hq_0 U^\dagger(\x).
\end{align}
This parameterization preserves the non-linear constraint $\hq^2 = \I_{4R}$. The generator $\hw$ describes fluctuations around the causal saddle point $\hq_0 \propto \tau_z^{\rm K}$ [cf.~\cref{eq:ansatz_q0}]. Coset transversality $\{\hw, \hq_0\} = 0$ together with Class CI anti-unitary invariance restricts the parent-manifold generator to the Keldysh off-diagonal form $\hw = \tau_x^{\rm K} \otimes W_1$ with $W_1 \in \mathfrak{usp}(2R)$ [cf.~\cref{ap:G_hw}]. The parent manifold $\mathcal{M}_{\text{parent}} \simeq \USp(2R)$ has dimension $R(2R+1)$.

A finite measurement rate $\gamma > 0$ breaks this symmetry, inducing a mass gap for a subset of these fluctuations. This mass generation restricts the low-energy dynamics to a smaller massless submanifold. To isolate these surviving soft modes, we must evaluate the measurement action expanded to quadratic order in $\hw$ [cf.~\cref{eq:LM_W_2}].

In the rare-measurement limit, expanding the measurement action [cf.~\cref{eq:LM_det_strict}] to second order in $\hw$ (see \cref{ap:mea_f}) yields a local term. The linear contribution vanishes by the saddle-point condition, while the derivative-free quadratic term generates a mass for the fluctuations:
\begin{align}\label{eq:LM_W_2}
    \i \L_M^{(2)}[\hw] \sim - \sum_{\beta\in\{c,s\}} \tr\Bigl( [\hw, \hm_\beta]^2 \Bigr).
\end{align}
This quadratic measurement term introduces a mass gap for a subset of the fluctuations. If a generator commutes with the measurement matrix ($[\hw,\hm_\beta] = 0$), \cref{eq:LM_W_2} vanishes, leaving the mode massless. If it anticommutes ($\{\hw, \hm_\beta\} = 0$), the relation $[\hw, \hm_\beta] = 2\hw\hm_\beta$ and the normalization $\hm_\beta^2 = \I$ yield a mass term $\frac{1}{8}\tr(\hw^2)$ per channel. These massive modes decouple from the low-energy spectrum. Because the generator $\hw$ is anti-Hermitian, the trace obeys $\tr(\hw^2) \le 0$. Consequently, the negative sign in the Euclidean action $S_E \propto - M \int \tr(\hw^2)$ ensures a positive energy penalty for these fluctuations, confirming the positivity of the mass gap. In the Keldysh subspace, both measurement matrices $\hm_\beta$ are proportional to $\tau^{\rm K}_x$ [cf.~\cref{eq:hm12_def}]. Because $\{\tau^{\rm K}_y, \tau^{\rm K}_x\} = 0$, fluctuation components containing $\tau^{\rm K}_y$ acquire a mass gap, whereas components proportional to $\tau^{\rm K}_x$ remain gapless.

In the particle-hole subspace, the charge-channel matrix $\hm_c \propto \mu^{\rm PH}_z$ anticommutes with $\mu^{\rm PH}_x$ and $\mu^{\rm PH}_y$, generating a mass gap for these components. On the soft subspace parameterized by $\hw = \tau^{\rm K}_x \otimes W_1$, the spin-channel matrix $\hm_s \propto \mu^{\rm PH}_0$ satisfies $[\hw, \hm_s] = 0$ identically and yields no mass term. Thus, the modes that commute with both measurement channels assume the form $\hw \propto\tau^{\rm K}_x \otimes \mu^{\rm PH}_{\alpha}$, where $\alpha \in \{0, z\}$. These commutant generators span the unitary Lie algebra $\mathfrak{u}(R)$, reducing the dimension of the soft subspace from $R(2R+1)$ to $R^2$ and freezing $R(R+1)$ massive modes.

Next, we evaluate the pairing-induced mass $M_\Delta$ for the fluctuations in the $\mu_z^{\rm PH}$ channel. Expanding the trace-log action \cref{eq:S0_hq_final} to quadratic order in the fluctuation generator $\hw_z = \tau_x^{\rm K} \otimes \mu_z^{\rm PH} \otimes \mathcal{R}_z$ yields the second-order action (see \cref{ap:pairing_mass}):
\begin{align}\label{eq:S0_quad_mass}
    \i S_{0}^{(2)}[\hw_z] = \frac{M_\Delta}{2} \int \dd^2\x\, \tr(\hw_z^2).
\end{align}
Because the pairing term $-\Delta \mu_x^{\rm PH}$ anticommutes with $\mu_z^{\rm PH}$, the superconducting gap breaks the Ward cancellation that protects the Goldstone channel, generating the mass coefficient:
\begin{align}\label{eq:M_Delta_formula}
M_\Delta = \frac{\pi \nu_F \gamma \Delta^2}{2\sqrt{\Delta^2+\gamma^2}} \approx \frac{\pi}{2} \nu_F \gamma \Delta \quad (\gamma \ll \Delta),
\end{align}
where $\nu_F$ is the normal-state density of states. This finite mass gaps out the $\mu_z^{\rm PH}$ channel, removing it from the low-energy spectrum. Consequently, the surviving massless fluctuations are restricted to:
\begin{align}\label{eq:hw_surviving}
\hw(x,t) = \tau^{\rm K}_x \otimes \mu^{\rm PH}_0 \otimes \mathcal{R}(x,t),
\end{align}
where the matrix $\mathcal{R}$ acts on the replica space.

The geometry of the NLSM manifold is determined by two intrinsic constraints on $\hw$. First, generating unitary rotations requires $\hw^\dagger = -\hw$. Because the matrices $\tau^{\rm K}_x$ and $\mu^{\rm PH}_0$ are Hermitian, the replica matrix must be anti-Hermitian, $\mathcal{R}^\dagger = -\mathcal{R}$. Next, $\hw$ should belong to the Lie algebra of the parent Class CI group [cf.~\cref{eq:parent_symmetry_G}], satisfying the symplectic condition $\hw^{\rm T}\hx + \hx \hw = 0$ for the metric $\hx$ [cf.~\cref{ap:G_hw}]. Substituting the expression from \cref{eq:hw_surviving} into this condition, and using the symmetries $(\tau^{\rm K}_x)^{\rm T} = \tau^{\rm K}_x$ and $(\mu^{\rm PH}_0)^{\rm T} = \mu^{\rm PH}_0$, yields $\mathcal{R}^{\rm T} = -\mathcal{R}$.

Because $\mathcal{R}$ is anti-Hermitian ($\mathcal{R}^\dagger =-\mathcal{R}$) and antisymmetric ($\mathcal{R}^{\rm T} = -\mathcal{R}$), it is real. These real antisymmetric matrices span the $\mathfrak{so}(R)$ Lie algebra, establishing $\SO(R)$ as the NLSM manifold for the single $f$-fermion sector.
\begin{align}\label{eq:SO_R_manifold}
\mathcal{M}_{\text{NLSM}} = \SO(R).
\end{align}
Because the canonical transformation to the $f$-fermion basis accounts for all microscopic degrees of freedom without redundancy, this single $\SO(R)$ NLSM constitutes the complete low-energy effective field theory of the monitored system. The $\SO(R)$ manifold has dimension $R(R-1)/2$, which evaluates to zero at $R=1$. This vanishing dimension indicates that the $R=1$ limit represents a formal replica constraint rather than a literal dynamical target space. We construct the NLSM framework at generic integers $R \ge N$. Because the resulting beta function $\beta(g) = (R-2)g^2/16\pi$ [cf.~\cref{eq:beta_function_recap}] depends smoothly on $R$, we apply the $R \to 1$ limit algebraically through analytic continuation. This standard Keldysh procedure ensures a well-defined renormalization group flow independent of the vanishing geometric dimension.

\subsection{Topological Action and Temporal Dynamics}\label{sec:WZW}
A gradient expansion of the bare action $S_0[\hq]$ around the saddle point gives the effective action for the temporal dynamics. As detailed in \cref{ap:Stiffness}, expanding this action to first order in the time derivative yields the following kinetic term: \begin{align}\label{eq:S_kin_t1_corrected} 
\i S_{\kin}^{(1)}[\hq] \sim\nu_F Z_\omega \int \dd^2\x\, \tr(\hq_0 \hq \partial_t \hq), 
\end{align} 
where $\nu_F$ is the normal-state density of states and $Z_\omega = 2\gamma/\sqrt{\Delta^2+\gamma^2}$ is the dimensionless frequency renormalization factor [cf.~\cref{eq:I_t_1}]. The dynamical critical exponent $z$ characterizes the low-energy scaling through the dispersion relation $\omega \propto k^z$, obtained by comparing the scaling dimensions of the temporal and spatial derivative terms. In conventional disordered systems, such a first-order temporal derivative $\tr(\hq_0 \hq \partial_t \hq)$ provides a linear frequency dependence that balances against the spatial gradient $\tr(\partial_x \hq)^2 \sim k^2$, leading to diffusive dynamics with dynamical exponent $z=2$.

We evaluate the soft-mode contribution to the first-order temporal kinetic action [cf.~\cref{eq:S_kin_t1_corrected}] by substituting the unitary rotation $U(\x)$ into the equivalent trace structure $\tr(\hq_0 U^\dagger \partial_t U)$ derived in \cref{ap:Stiffness}. The $\SO(R)$ NLSM manifold confines the generator to the matrix form $\hw \propto \tau^{\rm K}_x \otimes \mu^{\rm PH}_0 \otimes \mathcal{R}$ [cf.~\cref{eq:hw_surviving}]. Expanding the trace to the leading order in the fluctuation generator produces $\tr(\hq_0 \partial_t \hw)$. Because the saddle point assumes the proportional structure $\hq_0 \propto \tau^{\rm K}_z$ [cf.~\cref{eq:ansatz_q0}], this leading term evaluates to zero as a consequence of the orthogonality between Pauli matrices in the Keldysh space: $\tr_K(\tau^{\rm K}_z \tau^{\rm K}_x) = 0$. As derived in \cref{ap:vanish}, this cancellation extends to all orders in $\hw$ because $\tr_K[\tau^{\rm K}_z (\tau^{\rm K}_x)^n] = 0$ for any integer $n$. This all-order vanishing of the first-order temporal derivative suppresses the diffusive $z=2$ channel on the projected $\SO(R)$ manifold, indicating that the low-energy dynamics are governed instead by second-order temporal derivatives.

The first-order temporal kinetic action $\i S_{\kin}^{(1)}[\hq]$ vanishes on the $\SO(R)$ manifold due to its algebraic structure, where the trace over the generators in the linear temporal sector evaluates to zero. We therefore examine whether a topological contribution can provide the leading temporal dynamics. Because the second homotopy group of any compact Lie group vanishes, $\pi_2(\SO(R)) = 0$, a topological $\theta$-term is mathematically prohibited on the $\SO(R)$ manifold. In contrast, the $\SO(R)$ manifold admits Wess--Zumino--Witten (WZW) terms~\cite{witten_non-abelian_1984} in principle, since $\pi_3(\SO(R)) \cong \mathbb{Z}$ for $R \ge 3$. A WZW action $S_{\text{WZW}}$ can be constructed by integrating a volume form over a three-dimensional auxiliary extension of the spacetime manifold. The matrix field $\hq(t,x)$ is extended into an auxiliary field $\tilde{\hq}(u, t, x)$ parameterized by $u \in [0, 1]$, with boundary conditions $\tilde{\hq}(0, t, x) = \hq_0$ and $\tilde{\hq}(1, t, x) = \hq(t, x)$. We construct the Wess-Zumino-Witten (WZW) term directly on the orthogonal group manifold $\SO(R)$. While the microscopic fermion theory belongs to Class CI, the effective low-energy target is $\SO(R)$, meaning its topological properties are dictated by the homotopy groups of $\SO(R)$. Under this construction, the WZW action takes the form:
\begin{align} \label{eq:WZW_general}
&S_{\text{WZW}}[\hq]
\\\sim& \i k \int_0^1 \dd u \int \dd^2\x \, \epsilon^{\mu\nu\lambda} \tr\left( \tilde{\hq}^{-1} (\partial_\mu \tilde{\hq})
\tilde{\hq}^{-1} (\partial_\nu \tilde{\hq}) \tilde{\hq}^{-1} (\partial_\lambda
\tilde{\hq}) \right),
\end{align}
where $\epsilon^{\mu\nu\lambda}$ ($\mu, \nu, \lambda \in \{u, t, x\}$) is the Levi-Civita symbol. The level $k$ is quantized, because the difference between any two smooth extensions of the field forms a map from $S^3$ to $\SO(R)$, and the single-valuedness of the path-integral weight $e^{\i S_{\text{WZW}}}$ restricts $k$ to integer values.

For the $s$-wave BCS Hamiltonian in one dimension, the topological classification for symmetry class CI is trivial (invariant 0)~\cite{schnyder_classification_2008,chiu_classification_2016}. The two-component BdG vector $\vec{d}(k) = (-\Delta, 0, \xi_k)$ maintains a non-zero component $d_x = -\Delta \ne 0$ throughout the Brillouin zone, yielding a vanishing winding number $W = 0$. Furthermore, the anti-unitary time-reversal symmetry $\widetilde{\mathcal T}$ pairs the Fermi points at $\pm k_F$ with opposite chiralities, canceling any non-Abelian chiral anomaly. Consequently, the quantized topological level is zero ($k = 0$), ruling out any WZW contribution ($S_{\text{WZW}} \equiv 0$). The leading temporal dynamics are governed by the second-order derivative terms derived below.

Because both the first-order temporal kinetic term and the WZW action vanish on the $\SO(R)$ manifold, the low-energy effective action is determined by the second-order dynamical terms evaluated at one-loop order (see \cref{ap:kin}, \cref{ap:second_t}):
\begin{align}\label{eq:S_kin_final}
\i S_\eff[\hq] \sim\nu_F\int \dd^2\x\, \tr \left[ -D_x (\partial_x \hq)^2 + D_t (\partial_t \hq)^2 \right].
\end{align}
While the single-particle excitation spectrum is gapped by $\Delta$, the low-energy orientational modes in Keldysh-replica space remain gapless Goldstone modes. As derived in \cref{ap:kin}, the stiffness coefficients evaluate to $D_t = \frac{J}{\gamma}$ and $D_x = \frac{v_F^2 \Delta^2}{2\gamma\sqrt{\Delta^2+\gamma^2}}$, where $v_F = |\partial_k \xi_k|_{k_F}$ is the Fermi velocity and $J$ denotes the hopping amplitude. In the regime $\gamma \ll \Delta$, the spatial stiffness reduces to $D_x \approx \frac{v_F^2 \Delta}{2\gamma}$, leading to the velocity $v_{\rm eff} = \sqrt{D_x/D_t} \approx \sqrt{2J\Delta}$. The rate $2\gamma$ defines the mean free path $l_0 \equiv v_F / (2\gamma)$, whose inverse $\Lambda = l_0^{-1}$ sets the ultraviolet momentum cutoff for the renormalization group analysis in \cref{sec:EE}. Balancing the derivative terms in \cref{eq:S_kin_final} yields the linear dispersion $\omega \propto k$ with dynamical critical exponent $z=1$ and effective spacetime dimension $D_{\text{eff}} = 2$.

\section{Renormalization Group Analysis}\label{sec:RG}
\subsection{Background Field Parameterization and Derivation of the Fluctuation Action}\label{sec:slow_fast}

The infrared properties of the effective theory are evaluated using the Wilsonian momentum-shell renormalization group (RG) framework~\cite{wilson_renormalization_1974,poboiko_PhysRevX.13.041046}. This approach coarse-grains the short-distance fluctuations by progressively integrating out high-momentum modes. We introduce an ultraviolet momentum cutoff $\Lambda$ to regularize the momentum integrations
\begin{align}\label{eq:Lambda_l_0}
\Lambda \equiv l_0^{-1} = \frac{2\gamma}{v_F}.
\end{align}
This cutoff defines the upper boundary for the modes included in the effective action. 

We formulate the theory in Euclidean space through a Wick rotation $t \to -\i \tau$. Defining the scaled imaginary time
\begin{align}\label{eq:x_0_tau}
x_0 \equiv \sqrt{\frac{D_x}{D_t}} \, \tau
\end{align}
absorbs the spatial-temporal anisotropy of the gradient terms in \cref{eq:S_kin_final} and establishes continuous rotational invariance in the isotropic coordinates $\bm{r} \equiv (x, x_0)$. Performing the Wick rotation $t \to -\i\tau$, evaluating the Keldysh and particle-hole trace $\tr_{4R}[(\nabla \hq)^2] = 4\tr[(\nabla Y)(\nabla Y^{\rm T})]$, and defining the dimensionless coupling $1/g \equiv \frac{\pi \nu_F}{2}\sqrt{D_x D_t}$ yields the isotropic Euclidean action:
\begin{align}\label{eq:euclidean_action}
  S_\eff[Y] \sim \frac{1}{g} \int \dd^2 \bm{r} \, \tr \left[ (\nabla Y) (\nabla Y^{\rm T}) \right] = -\frac{1}{g} \int \dd^2 \bm{r} \, \tr (J_\mu^2),
\end{align}
where $J_\mu = Y^{\rm T}\partial_\mu Y \in \mathfrak{so}(R)$ is the Lie algebra current. In the regime $\gamma \ll \Delta$, evaluating the stiffness coefficients yields $\sqrt{D_x D_t} \approx \frac{v_F \sqrt{J\Delta}}{\sqrt{2}\gamma}$, which determines the dimensionless bare coupling as:
\begin{align}\label{eq:bare_coupling_def}
    \frac{1}{g_0} \approx \frac{\pi \nu_F}{2}\sqrt{D_x D_t} = \frac{\sqrt{J\Delta}}{2\sqrt{2}\gamma},
\end{align}
where $\nu_F = 1/(\pi v_F)$ is the density of states at half filling. The bare stiffness $1/g_0$ is governed by the ratio of the hybrid energy scale $\sqrt{J\Delta}$ to the measurement dephasing rate $\gamma$. The one-loop weak-coupling expansion is controlled when $g_0 \sim \frac{\gamma}{\sqrt{J\Delta}} \ll 1$, which is satisfied in the regime $\gamma \ll \sqrt{J\Delta} \ll J$. Beyond this strict analytical window where $g_0 \gtrsim 1$, the numerical simulations in the companion Letter~\cite{guo2026} are consistent with the persistence of the super-logarithmic scaling.

This emergent rotational invariance in the $(x, x_0)$ plane is a consequence of the dynamical exponent $z=1$ derived in \cref{sec:WZW}, placing spatial and temporal gradients on equal footing in the infrared. The low-energy rotational degrees of freedom on the $\SO(R)$ manifold are parameterized by the Euclidean field $Y(\bm{r})$. To evaluate the RG flow, we decompose the field via the background-fluctuation map $Y(\bm{r}) = Y_0(\bm{r}) \exp[\mw(\bm{r})]$, where $Y_0(\bm{r}) \in \SO(R)$ contains slow modes with momenta $0 \le |\q| < \Lambda/b$ and $\mw(\bm{r}) \in \mathfrak{so}(R)$ denotes fast fluctuations within the shell $\Lambda/b \le |\q| < \Lambda$ with $b = e^{\dd\ell}$.

Integrating out the fast modes requires expanding the action to second order in $\mw$. As derived in \cref{ap:momentumshell}, the curvature mass terms cancel algebraically, preserving the gapless nature of the soft modes. Evaluating the momentum-shell contractions yields the one-loop quantum correction:
\begin{align}\label{eq:Strict_Shell_Result}
\Delta S^{\text{RG}} \sim - \left[ c_\beta(R-2) \ln b \right] \int \dd^2\bm{r}\, \tr \left[ (\nabla Y_0) (\nabla Y_0^{\rm T}) \right],
\end{align}
where $c_\beta > 0$ is a positive momentum-shell coefficient. Rescaling the spatial coordinates $\bm{r} \to \bm{r}/b$ restores the ultraviolet cutoff $\Lambda$ and determines the scale dependence of the running coupling.

\subsection{Scaling form of Entanglement Entropy}\label{sec:EE}
Adding the one-loop correction in \cref{eq:Strict_Shell_Result} to the bare action in \cref{eq:euclidean_action} determines the running coupling $g(b)$ at the rescaled cutoff $\Lambda/b$:
\begin{align}\label{eq:running_coupling_def}
\frac{1}{g(b)} \approx \frac{1}{g} - c_\beta (R-2) \ln b,
\end{align}
where $c_\beta$ is the one-loop momentum-shell coefficient determined by \cref{eq:Strict_Shell_Result}. The RG flow is parameterized by the logarithmic scale $\ell \equiv \ln b = \ln(r/l_0)$. Differentiating \cref{eq:running_coupling_def} with respect to $\ell$ yields the one-loop beta function:
\begin{align}\label{eq:beta_function_recap}
\beta(g) \equiv \frac{\dd g}{\dd \ell} = c_\beta (R-2) g^2 ,\quad \beta(g)\bigg|_{R \to 1} = -c_\beta g^2.
\end{align}
In the replica limit $R \to 1$, the negative beta function drives the coupling toward the weak-coupling limit $g \to 0$ in the infrared. Integrating \cref{eq:beta_function_recap} from the bare condition $g(0) = g_0$ gives the running stiffness:
\begin{align}\label{eq:explict_flow}
\frac{1}{g(\ell)} = \frac{1}{g_0} + c_\beta \ell.
\end{align}
The scale-dependent effective coupling decreases as $g(\ell) \approx 1/(c_\beta \ell)$ in the infrared limit. This negative flow generates a weak-anti-localization effect, where the effective stiffness increases with length scale and prevents localization into an area-law state.

We evaluate the scaling behavior of the entanglement entropy by mapping the ensemble-averaged density matrix $\overline{\Tr(\rho_A^N)}$ to the expectation value of a cyclic permutation operator. In the replica framework, this operator inserts a pair of twist defects, $\mathcal{T}_N$ and $\tilde{\mathcal{T}}_N$, at the boundaries of subsystem $A$ of length $L$~\cite{Pasquale_Calabrese_2004, Fava_PhysRevX.13.041045}. The von Neumann entanglement entropy $\se(L)$ is extracted from the twist-defect free energy $\Delta F_N(L) = -\ln \langle \mathcal{T}_N(L) \tilde{\mathcal{T}}_N(0) \rangle$ through the replica derivative:
\begin{align}\label{eq:S_ee_limit}
\se(L) \sim -\lim_{N \to 1} \frac{\partial}{\partial N} \ln \langle \mathcal{T}_N(L) \tilde{\mathcal{T}}_N(0) \rangle.
\end{align}
We use the polar coordinates $(r, \theta)$ center on a twist defect, the boundary condition imposes the monodromy constraint $Y(r, \theta + 2\pi) = \Omega_N Y(r, \theta)$, where $\Omega_N \in \SO(N)$ denotes the cyclic permutation matrix for odd integers $N$. Minimizing the action under this constraint yields a classical vortex configuration $Y_{\text{cl}}(r, \theta) \sim \exp(\theta \hat{X})$, where $\hat{X} \in \mathfrak{so}(N)$ is the Lie algebra generator satisfying $\exp(2\pi \hat{X}) = \Omega_N$. Because the twist-defect pair at the boundaries of subsystem $A$ forms a vortex dipole whose far field is screened for $r \gg L$, the logarithmic integral is truncated at the separation $L$. Integrating this energy density over the spatial range $l_0 \le r \le L$ with the running stiffness yields an expression in powers of $\ln(L/l_0)$ [cf.~\cref{ap:EE}]. Expanding $\ln(L/l_0) = \ln L - \ln l_0$ preserves the universal leading coefficient $A \sim c_\beta \sim \mathcal{O}(1)$ and absorbs the cutoff scale $l_0 = v_F/(2\gamma)$ into the subleading contributions, yielding the scaling form:
\begin{align}\label{eq:entropy_microscopic_expanded}
\se(L) = A \ln^2 L + B \ln L + \text{const},
\end{align}
where $A > 0$ is a universal coefficient governed by the $\SO(R)$ weak-anti-localization flow, while $B \sim 1/g_0$ denotes the non-universal slope determined by the bare stiffness. The asymptotic regime dominated by the $\ln^2 L$ dependence emerges when the subsystem size exceeds the crossover scale $L_{\text{cross}} \sim l_0 \exp(B/A)$ [cf.~\cref{eq:L_cross_estimate}]. For system sizes accessible in numerical simulations, subsystem lengths lie within the crossover window $l_0 \ll L \lesssim L_{\text{cross}}$, where the subleading logarithmic term $B \ln L$ contributes substantially to the effective finite-size slope extracted from numerical fits~\cite{guo2026}.

\section{Conclusion and Discussion}
\label{sec:conclusion}
In summary, we have developed a Keldysh-replica field theory for the non-equilibrium entanglement dynamics of a monitored spinful $s$-wave BCS chain. In the rare-measurement regime, $\gamma \ll J,\Delta$, the combined effect of pairing and measurement backaction dynamically projects the low-energy fluctuation sector onto an $\SO(R)$ target manifold in replica space. The one-loop renormalization group flow of this $\SO(R)$ NLSM has a negative beta function in the replica limit $R\to1$, leading to weak anti-localization and the super-logarithmic entanglement scaling
$S_s(L)\sim \ln^2 L$.
This provides an analytical explanation of the measurement-enhanced entanglement scaling observed in the companion Letter~\cite{guo2026}.

The field theory is perturbatively controlled within the weak-coupling window $\gamma \ll \sqrt{J\Delta} \ll J$, where $g_0 \ll 1$. When the measurement rate becomes comparable to the bandwidth or pairing gap, $\gamma \gtrsim J,\Delta$, the gradient expansion underlying the NLSM breaks down. In this strong-measurement regime, quantum Zeno localization suppresses long-range entanglement and drives the system toward an area-law phase. A possible mechanism for this localization transition involves the proliferation of $\mathbb{Z}_2$ topological vortices. In the weak-coupling phase, such vortices remain energetically confined, preserving the weak-anti-localizing scaling. At stronger measurements, vortex entropy can overcome the action cost, leading to vortex deconfinement and localization. Because the $s$-wave model lacks a WZW term, topological destructive interference between vortex configurations is absent, and the transition is unobstructed by static topological invariants. A quantitative treatment of this strong-measurement crossover lies beyond the one-loop continuum field theory.

A central result of our analysis is that the $\SO(R)$ target does not arise from Majorana degrees of freedom. The monitored spinful $s$-wave BCS system belongs to symmetry class CI, with a parent symplectic saddle manifold $\mathcal M_{\rm parent} \simeq \USp(2R)$. The measurement and pairing terms impose complementary mass constraints on this parent manifold: the charge measurement term selects the commutant $\U(R)$ subgroup by gapping out the $\mu_{x}^{\rm PH}$ and $\mu_{y}^{\rm PH}$ particle-hole channels, while the pairing term removes the remaining $\mu_z^{\rm PH}$ channel. After these massive modes are integrated out, the symplectic constraint forces the surviving replica generators to be real and antisymmetric, yielding the effective Lie algebra $\mathfrak{so}(R)$ and the soft-mode target manifold $\SO(R)$.

This mechanism distinguishes the monitored $s$-wave BCS chain from previously studied topological Majorana $p$-wave systems. In the latter, the orthogonal structure is tied more directly to Majorana variables and may be accompanied by a nonzero WZW term. In contrast, the present model is topologically trivial and has no WZW contribution. Nevertheless, it exhibits the same perturbative $\SO(R)$ weak-anti-localizing flow and the same asymptotic $S_s(L)\sim \ln^2 L$ scaling. Our results therefore show that measurement-induced super-logarithmic criticality does not require static topological protection; it can emerge dynamically from the interplay of superconducting pairing, measurement backaction, and replica-space soft modes.

\textit{Acknowledgement.}--
We thank Ji-Yao Chen for insightful discussions. RJG is supported by National Natural Science Foundation of China (Grants No. 12447107, No. 12304186), Guangdong Basic and Applied Basic Research Foundation (Grant No. 2024A1515013065), and Quantum Science and Technology - National Science and Technology Major Project (Grant No. 2021ZD0302100).

\bibliography{ref}
\appendix
\renewcommand{\theequation}{\Alph{section}\arabic{equation}}
\setcounter{equation}{0}
\crefalias{section}{appendix}
\crefalias{subsection}{appendix}

\section{Keldysh Rotation and Measurement Determinant}\label{ap:Keldysh}

This appendix supplies the technical steps used in
\cref{sec:keldysh_formalism}. We first define the Keldysh contour and its observation-time boundary. We then perform the LO rotation and derive the equal-time regularization of the projection operators. Next, we sum the measurement outcomes and obtain the charge and spin channels. Finally, we introduce the generalized Hubbard--Stratonovich fields and derive the all-channel determinant in the slow sector.

\subsection{Contour and observation-time boundary}

For each replica $r$, the Keldysh contour
\begin{align}
    \mathcal C_r=\mathcal C_{+,r}\cup\mathcal C_{-,r}
\end{align}
runs from the initial time $t_i$ to the observation time $t_f$ on the forward branch and returns from $t_f$ to $t_i$ on the backward branch. The observable or replica twist is inserted at the turning time $t_f$.

For a fixed measurement history $\mathcal M=\{(j_m,t_m,\mathbf n_m)\}_{m=1}^{M}$ with $t_i\leq t_m\leq t_f$, the event positions, times, and outcomes act identically in every replica. The corresponding projector insertion has the contour-ordered form
\begin{align}\label{eq:fixed_measurement_history}
    \mathcal W_{\mathcal M}=\mathcal T_{\mathcal C}\prod_{m=1}^{M}\prod_{r=1}^{R} \hat P_{\mathbf n_m}^{+,r}(j_m,t_m)\hat P_{\mathbf n_m}^{-,r}(j_m,t_m).
\end{align}
Averaging over the Poisson distribution of event numbers and positions gives
\begin{align}
    \overline{\mathcal W_{\mathcal M}}_{\rm kin}
    &=
    \sum_{M=0}^{\infty}\frac{e^{-\overline{M}}\overline{M}^M}{M!}
    \nn
    \quad\times\prod_{m=1}^{M}\left[\sum_{\mathbf n_m}\sum_{j_m=1}^{L}\int_{t_i}^{t_f}\frac{\dd t_m}{TL}\,\prod_{r=1}^{R}(\hv_{\mathbf n_m})_{+,r}(\hv_{\mathbf n_m})_{-,r}\right]
    \en
    e^{-\overline{M}}\exp\!\left[\gamma\int_{t_i}^{t_f}\dd t\int\dd x\,\sum_{\mathbf n}\prod_{r=1}^{R}(\hv_{\mathbf n})_{+,r}(\hv_{\mathbf n})_{-,r}\right]
    \en
    \exp\!\left(\i\gamma S_M\right),
\end{align}
where $\overline{M}=\gamma L(t_f-t_i)=\gamma\int_{t_i}^{t_f}\dd t\int\dd x$, and the measurement action is
\begin{align}\label{eq:bounded_measurement_action_appendix}
    S_M=\int_{-\infty}^{\infty}\dd t\int\dd x\,\Theta(t-t_i)\Theta(t_f-t)\L_M.
\end{align}
Thus, the term $-1$ in $\L_M$ originates from the Poisson no-event factor $e^{-\overline{M}}$.

The role of the upper boundary can be seen from the outcome sum. Define
\begin{align}
    X_{\mathbf n,r}
    \equiv
    (\hv_{\mathbf n})_{+,r}(\hv_{\mathbf n})_{-,r}.
\end{align}
For one replica, the completeness and idempotency of the projectors restore the normalized closed-contour evolution after summing over $\mathbf n$. For several replicas, however, the shared outcome produces
\begin{align}\label{eq:replica_outcome_nonfactorization}
    \sum_{\mathbf n}\prod_{r=1}^{R}X_{\mathbf n,r}\neq\prod_{r=1}^{R}\left(\sum_{\mathbf n}X_{\mathbf n,r}\right), \quad R\neq1.
\end{align}
The measurement average therefore couples the replicas and cannot be reduced to independent single-replica Keldysh normalizations. Extending the measurement term beyond $t_f$ would add shared measurement events that are absent from the replicated observable.

One may extend the unitary parts of the forward and backward contours beyond $t_f$ for formal purposes, provided that $\gamma(t>t_f)=0$. The remaining unitary evolution then cancels separately within each replica. For the stationary bulk saddle, the duration $T=t_f-t_i$ is taken to infinity, and the fields are evaluated far from both temporal endpoints. This limit permits the frequency representation used in the main text while omitting endpoint contributions.

\subsection{LO rotation and equal-time regularization}

For a local outcome $\mathbf n=(n_\uparrow,n_\downarrow)$, the projector symbols on either Keldysh branch are
\begin{subequations}\label{eq:proj_ops}
\begin{align}
    (\hv_{(0,0)})_{\pm,r}
    &=\left(1-\bp_\uparrow\p_\uparrow \right)_{\pm,r}
    \left(1-\bp_\downarrow\p_\downarrow\right)_{\pm,r},
    \label{eq:P1}
    \\
    (\hv_{(0,1)})_{\pm,r}
    &=\left(1-\bp_\uparrow\p_\uparrow \right)_{\pm,r}
    \left(\bp_\downarrow\p_\downarrow\right)_{\pm,r},
    \label{eq:P2}
    \\
    (\hv_{(1,0)})_{\pm,r}
    &=\left(\bp_\uparrow\p_\uparrow \right)_{\pm,r}
    \left(1-\bp_\downarrow\p_\downarrow\right)_{\pm,r},
    \label{eq:P3}
    \\
    (\hv_{(1,1)})_{\pm,r}
    &=\left(\bp_\uparrow\p_\uparrow \right)_{\pm,r}
    \left(\bp_\downarrow\p_\downarrow\right)_{\pm,r}.
    \label{eq:P4}
\end{align}
\end{subequations}
The two spin factors can be treated separately before summing the four joint outcomes.
 Let $\alpha\in\{1,2\}$ denote the particle-hole component.

\paragraph{Canonical particle-hole map.}
The transformed operators satisfy canonical anticommutation relations $\{f_{j,\alpha}, f_{j',\beta}^\dagger\} = \delta_{jj'}\delta_{\alpha\beta}$. In terms of the original electron densities, the transformed local occupation numbers are
\begin{align}
    f_{j,1}^\dagger f_{j,1} = \hat n_{j,\uparrow}, \quad
    f_{j,2}^\dagger f_{j,2} = c_{j,\downarrow}c_{j,\downarrow}^\dagger = 1-\hat n_{j,\downarrow}.
\end{align}
Summing these local contributions gives
\begin{align}\label{eq:Nfj_Sz_appendix}
    N_{f,j} \equiv f_{j,1}^\dagger f_{j,1}+f_{j,2}^\dagger f_{j,2} = 1+2S_{z,j}.
\end{align}
Summing over all $L$ lattice sites confirms the global identity
\begin{align}\label{eq:Nf_Sz_appendix}
    N_f \equiv \sum_{j=1}^{L}N_{f,j} = L+2S_z, \quad S_z \equiv \sum_{j=1}^{L}S_{z,j}.
\end{align}
We apply the Larkin-Ovchinnikov rotation
\begin{subequations}\label{eq:ALVO}
    \begin{align}
        \p_1&=\frac{\p_+ + \p_-}{\sqrt{2}},
        \quad
        \p_2=\frac{\p_+ - \p_-}{\sqrt{2}},
        \\
        \bp_1&=\frac{\bp_+ - \bp_-}{\sqrt{2}},
        \quad
        \bp_2=\frac{\bp_+ + \bp_-}{\sqrt{2}}.
    \end{align}
\end{subequations}
The branch-symmetric density then becomes
\begin{align}
    \bp_+\p_++ \bp_-\p_-= \bp_1\p_2+ \bp_2\p_1,
\end{align}
which fixes the $\tau_x^{\rm K}$ structure of the measurement vertices.

\paragraph{Composite fields and Green-function convention.}
We combine the LO, particle-hole, and replica labels into the composite index
\begin{align}\label{eq:composite_field_appendix}
    a=(b,\alpha,r),
    \quad
    \P_a=(\P_{b,r})_\alpha,
    \quad
    \BP_a=(\BP_{b,r})_\alpha.
\end{align}
For a fixed replica $r$, the fields $\P_r$ and $\BP_r$
contain four Grassmann components. The full fields $\P$ and $\BP$ contain $4R$ components and follow the tensor-product order Keldysh $\otimes$ particle-hole $\otimes$ replica. The identity matrices in these spaces are
\begin{align}
    \I_4=\tau_0^{\rm K}\otimes\mu_0^{\rm PH},
    \quad
    \I=\I_4\otimes\I_R.
\end{align}
For a matrix $A$ acting in the single-replica Keldysh
$\otimes$ particle-hole space,
\begin{align}\label{eq:replica_lift_identity}
    \BP(A\otimes\I_R)\P
    =\sum_{r=1}^{R}\BP_r A\P_r.
\end{align}
For the spacetime labels $1=(j,t_1)$ and $2=(j',t_2)$, our Green-function convention is
\begin{align}\label{eq:Green_function_convention}
    G_{ab}(1,2)
    =-\i\left\langle\P_a(1)\BP_b(2)\right\rangle.
\end{align}
The barred fields in this definition are independent
Grassmann variables.

\paragraph{Initial distribution.}
For the stationary reference state, $f_0(\omega)$ denotes the occupation of a single-particle mode at energy $\omega$, and
\begin{align}\label{eq:f0_F0_appendix}
    F_0(\omega)=1-2f_0(\omega)
\end{align}
is the associated Keldysh distribution function. The values $f_0=0$, $f_0=1$, and $f_0=1/2$ correspond to $F_0=1$, $F_0=-1$, and $F_0=0$, respectively.

The temporal-boundary data of a BCS state can, in general, be represented by a Hermitian distribution matrix in particle-hole space,
\begin{align}\label{eq:general_distribution_matrix}    \widehat F_0(k,\omega) = \sum_{\nu\in\{0,x,y,z\}}F_{0,\nu}(k,\omega)\mu_\nu^{\rm PH},
\end{align}
where the coefficients $F_{0,\nu}(k,\omega)$ are real. The components proportional to $\mu_x^{\rm PH}$ and $\mu_y^{\rm PH}$ describe anomalous distribution coherences, while the component proportional to $\mu_z^{\rm PH}$ describes a particle-hole imbalance. A general Hermitian matrix of this form is boundary distribution data; for a stationary state of the quadratic dynamics generated by $h_f$, it must in addition obey $[\,\widehat F_0(k,\omega),h_f(k)\,]=0$.

The reference distribution used in the main text is restricted to
\begin{align}\label{eq:scalar_reference_distribution}
    \widehat F_0(k,\omega)=F_0(\omega)\mu_0^{\rm PH}.
\end{align}
This choice removes the momentum dependence and sets the particle-hole imbalance and anomalous distribution components to zero. It does not remove the anomalous propagation generated by the pairing term in \cref{eq:Hamiltonian_k}.

\paragraph{Half filling, the reference regulator, and the saddle distribution.}
For a normalized single-particle density of states $\nu(\omega)$, half filling imposes the integrated condition
\begin{align}\label{eq:half_filling_integrated_condition}
    \int\dd\omega\,\nu(\omega) \left[f_0(\omega)-\frac{1}{2}\right]=0,
    \quad
    \int\dd\omega\,\nu(\omega)=1.
\end{align}
This condition allows a nonuniform occupation function. By contrast, 
\begin{align}\label{eq:F0_zero_pointwise_condition}     
F_0(\omega)=0 \quad\Longleftrightarrow\quad f_0(\omega)=\frac{1}{2} 
\end{align} 
for every $\omega$. This condition describes equal probabilities for single-particle mode occupations within the conserved symmetry sector. Because total $S_z$ and fermion parity are conserved by both Hamiltonian and measurement terms, an initial state with $S_z=0$ (such as the N\'eel state) remains confined to that subspace. Within this sector, the late-time unconditioned density matrix becomes maximally mixed, yielding the mode occupation $f_0(\omega)=1/2$. In the long-time bulk calculation, boundary transients decay, and $F_0(\omega)=0$ serves as the causal reference regulator for the bulk saddle point.

The bare function $F_0$ records temporal-boundary data associated with the chosen initial state. An actual vacuum or N\'eel initial condition need not satisfy $F_0(\omega)=0$ and may not be represented by the scalar stationary ansatz used for the bulk reference state. In the long-time bulk calculation, the corresponding boundary transients are omitted, and $F_0(\omega)=0$ is used as a convenient causal reference regulator.

For particle-hole flavor $\alpha\in\{1,2\}$ and replica $r$, define
\begin{align}
    x_{\alpha r}&\equiv\bar f_{\alpha,2,r}f_{\alpha,1,r}+\bar f_{\alpha,1,r}f_{\alpha,2,r},
    \\
    y_{\alpha r}&\equiv\bar f_{\alpha,2,r}f_{\alpha,1,r}\bar f_{\alpha,1,r}f_{\alpha,2,r}.
\end{align}

\paragraph{Equal-time prescription.}
For a propagator $G(t_1,t_2)$ depending on two time arguments, the regularised equal-time value is defined by the symmetric two-sided limit
\begin{align}\label{eq:G_reg_def}
    G^{\reg}(t,t)\equiv\frac{1}{2}\lim_{\varepsilon_t\to0^+}
    \bigg[&
        G\left(t+\frac{\varepsilon_t}{2}, t-\frac{\varepsilon_t}{2}\right)
        \nn+
        G\left(t-\frac{\varepsilon_t}{2}, t+\frac{\varepsilon_t}{2}\right)
    \bigg].
\end{align}
The limit is taken componentwise in Keldysh, particle-hole, and replica space. In a translation-invariant bulk, time-translation invariance gives
\begin{align}\label{eq:G_reg_stationary_appendix}
    G^{\reg}(0)=\frac{1}{2}\lim_{\varepsilon_t\to0^+}\left[G(\varepsilon_t)+G(-\varepsilon_t)\right],
\end{align}
where $G(t_1,t_2)=G(t_1-t_2)$.
The second relation is restricted to the stationary bulk, whereas near temporal boundaries the two-time definition in \cref{eq:G_reg_def} applies. The same prescription applies to the bare propagator $G_0$ and to dressed propagators $G$.

For a given particle-hole component $\alpha$ and replica $r$, we define the regularised branch-pair vertex by
\begin{align}
    V_{n,\alpha r}^{\reg}
    \equiv
    \left[
        (\hv_n)_{+,\alpha r}(\hv_n)_{-,\alpha r}
    \right]_{\rm reg},
    \quad
    n\in\{0,1\},
\end{align}
where $(\hv_n)_{\pm,\alpha r}$ are the single-branch symbols defined in \cref{eq:proj_ops}. The normal-ordered propagator and the symmetric propagator differ at coincident contour times by the step discontinuity of contour ordering. Under the LO rotation \eqref{eq:ALVO}, this discontinuity produces the local contact kernel
\begin{align}\label{eq:contact_contraction_rule}
    \i G_{ab}(j,t;j',t')&=\i G^{\reg}_{ab}(j,t;j',t')+\mathcal{C}_{ab}(j,t;j',t'),
    \\
    \mathcal{C}_{ab}(j,t;j',t')
    &\equiv
    -\frac{1}{2}\delta_{jj'}\delta_{tt'}
    \left( \tau_x^{\rm K}\otimes\mu_0^{\rm PH}\otimes\I_R\right)_{ab},
\end{align}
where $\delta_{tt'}$ is the Kronecker delta on the discretized time lattice.

We evaluate the counterterms by summing all partial Wick contractions with the contact kernel $\mathcal{C}$. For the empty-state projector pair $n_\alpha^{(f)}=0$, the normal-ordered product is
\begin{align}
    (\hv_0)_{+,\alpha r}(\hv_0)_{-,\alpha r}
    =
     &- \bigg(\bp_{\alpha,+}\p_{\alpha,+} 
    + \bp_{\alpha,-}\p_{\alpha,-}\bigg)_r
    \nn+
    \bigg(\bp_{\alpha,+}\p_{\alpha,+}\bp_{\alpha,-}\p_{\alpha,-}\bigg)_r+1.
\end{align}
Zero contact contractions preserve the quartic term, which equals $q_{\alpha r}$ in the LO basis. One contact contraction on the quartic term yields a bilinear contribution that combines with the bare bilinear to give $-\frac{1}{2}x_{\alpha r}$. Two contact contractions on the quartic term together with one contact contraction on the bilinear term yield the constant shift $\frac{1}{4}$. Repeating this contraction algebra for $n_\alpha^{(f)}=1$ gives
\begin{subequations}\label{eq:proj_ops_reg}
\begin{align}
    V_{0,\alpha r}^{\reg}=\frac{1}{4}-\frac{1}{2}x_{\alpha r}+y_{\alpha r},
    \quad
    V_{1,\alpha r}^{\reg}=\frac{1}{4}+\frac{1}{2}x_{\alpha r}+y_{\alpha r}.
\end{align}
\end{subequations}
The Grassmann algebra gives
\begin{align}
    x_{\alpha r}^2=2y_{\alpha r},
    \quad
    x_{\alpha r}^3=0.
\end{align}
For $\eta=\pm1$,
\begin{align}
    e^{2\eta x_{\alpha r}}=1+2\eta x_{\alpha r}+2x_{\alpha r}^2=1+2\eta x_{\alpha r}+4y_{\alpha r}.
\end{align}
Consequently, the two projectors for each particle-hole flavor can be written as
\begin{align}\label{eq:projector_reexp}
    V_{\eta,\alpha r}^{\reg}=\frac{1}{4}e^{2\eta x_{\alpha r}},
    \quad
    \eta=
    \begin{cases}
        -1,& n_\alpha^{(f)}=0,\\
        +1,& n_\alpha^{(f)}=1,
    \end{cases}
\end{align}
where $n_\alpha^{(f)}$ denotes the occupation of the fermion $f_\alpha$. These occupations are related to the original spin-resolved measurement outcomes by
\begin{align}
    n_1^{(f)}=n_\uparrow,
    \quad
    n_2^{(f)} = 1-n_\downarrow.
\end{align}
The corresponding signs for the original measurement outcomes are therefore
\begin{align}
    \eta_1=2n_\uparrow-1,
    \quad
    \eta_2=1-2n_\downarrow.
\end{align}

\subsection{Outcome sum and charge-spin channels}

Define $X_\alpha\equiv\sum_{r=1}^{R}x_{\alpha r}$.
For the two particle-hole flavors, \cref{eq:projector_reexp} gives
\begin{align}
    \prod_{r=1}^{R}
    V_{\eta_1,1r}^{\reg}
    V_{\eta_2,2r}^{\reg}
    =
    4^{-2R}
    \exp\!\left(
    2\eta_1X_1
    +
    2\eta_2X_2
    \right).
\end{align}
The sum over the four joint outcomes is therefore
\begin{align}\label{eq:outcome_sum_appendix}
    &\i\L_M[\BP,\P]+1
    \en
    4^{-2R}
    \sum_{\eta_1,\eta_2=\pm1}
    \exp\!\left(
    2\eta_1X_1
    +
    2\eta_2X_2
    \right)
    \en
    \frac{2}{4^{2R}}
    \left[
    \cosh\!\left(
    2X_1+2X_2
    \right)
    +
    \cosh\!\left(
    2X_1-2X_2
    \right)
    \right].
\end{align}
The two linear combinations correspond to the charge and spin measurement channels and satisfy
\begin{align}
    X_1-X_2= \BP\hm_c\P,
    \quad
    X_1+X_2=\BP\hm_s\P,
\end{align}
where the contour-symmetric matrices are
\begin{align}
    \hm_c= \tau_x^{\rm K}\otimes\mu_z^{\rm PH}\otimes\I_R,
    \quad
    \hm_s=\tau_x^{\rm K}\otimes\mu_0^{\rm PH}\otimes\I_R.
\end{align}
The constant shifts arising from the particle-hole transformation $c_\downarrow c_\downarrow^\dagger=1-\hat n_\downarrow$ are absorbed into the regularised exponent normalization $4^{-2R}$, leaving bilinear fermionic forms in the arguments. Substitution into \cref{eq:outcome_sum_appendix} yields
\begin{align}
    \i\L_M[\BP,\P]
    = \frac{2}{4^{2R}} \sum_{\beta\in\{c,s\}} \cosh\!\left( 2\BP\hm_\beta\P \right)-1,
\end{align}
which matches \cref{eq:L_M_final_res_strict}.

\subsection{Generalized Hubbard--Stratonovich representation
and all-channel determinant}\label{ap:GHS}

To formulate the Hubbard--Stratonovich representation, we begin with the convergent Gaussian identity
\begin{align}\label{eq:HS_convergent_appendix}
    1=\mathcal{N}_\epsilon\int\D\hg\int\D\hs\,\exp\!\left[-\frac{1}{2\epsilon}\Tr\!\left(\hg+\i\P\BP\right)^2-\frac{\epsilon}{2}\Tr\!\left(\hs^2\right)\right],
\end{align}
where $\epsilon>0$ regulates the integrals. Shifting $\hs\to\hs-\frac{\i}{\epsilon}(\hg+\i\P\BP)$ and taking $\epsilon\to0^+$ enforces $\hg=-\i\P\BP$ as a delta constraint. Reordering the Grassmann variables gives
\begin{align}
    \Tr(\hs\P\BP)=\sum_{a,b}\hs_{ab}\P_b\BP_a=-\sum_{a,b}\BP_a\hs_{ab}\P_b=-\BP\hs\P.
\end{align}
The total exponent in the functional integral is then
\begin{align}
    &\i\BP G_0^{-1}\P-\i\Tr\!\left[\hs\left(\hg+\i\P\BP\right)\right] \nn=\i\BP\left(G_0^{-1}+\i\hs\right)\P-\i\Tr(\hs\hg).
\end{align}
Integrating over the Grassmann fields yields the functional determinant
\begin{align}\label{eq:fermion_integration}&\int\D\BP\,\D\P\,\exp\!\left[\i\BP\left(G_0^{-1}+\i\hs\right)\P\right]
\nn=\exp\!\left[\Tr\ln\left(G_0^{-1}+\i\hs\right)\right].
\end{align}
Here, $\Tr\ln(\cdots)$ denotes the functional trace over spacetime coordinates as well as internal Keldysh, particle-hole, and replica indices, distinct from the local matrix determinant $\det_{4R}(\cdots)$ acting on internal indices alone.

To justify the all-channel replacement, we first place each local fermionic monomial in a fixed Grassmann order. A generic term containing $2m$ fields can be written as
\begin{align}\label{eq:generic_local_2N_vertex}
    \mathcal V_{2m}[\BP,\P]\equiv\int\dd^2\x\,\P_{a_1}(\x)\cdots\P_{a_m}(\x)\BP_{b_1}(\x)\cdots\BP_{b_m}(\x),
\end{align}
up to the fixed sign and matrix coefficients produced when the original measurement vertex is reordered into this form.

Let $K_i=(\omega_i,k_i)$ and $K_i'=(\omega_i',k_i')$ denote the frequency and momentum variables carried by $\P_{a_i}$ and $\BP_{b_i}$, respectively. The Fourier representation of the vertex contains
\begin{align}\label{eq:generic_vertex_momentum}
    \mathcal V_{2m}=&\int_{\{K_i,K_i'\}}(2\pi)^2\delta^{(2)}\left(\sum_{i=1}^{m}K_i-\sum_{i=1}^{m}K_i'\right)
    \nn\times\P_{a_1}(K_1)\cdots\P_{a_m}(K_m)\BP_{b_1}(K_1')\cdots\BP_{b_m}(K_m').
\end{align}
Here, $\int_{\{K_i,K_i'\}}$ denotes the product of all internal frequency and Brillouin-zone momentum integrations.

We define the slow-transfer projector by
\begin{align}\label{eq:appendix_slow_projector}
    \chi_\Lambda(Q)
    \equiv
    \Theta\!\left(\Lambda_\omega-|\Omega|\right)
    \Theta\!\left(\Lambda_q-|q|\right),
    \quad
    Q=(\Omega,q).
\end{align}
The two cutoff ratios are summarized by
\begin{align}\label{eq:slow_cutoff_ratio}
    \lambda
    \equiv
    \max\!\left(
        \frac{\Lambda_\omega}{E_{\rm uv}},
        \frac{\Lambda_q}{k_{\rm uv}}
    \right)
    \ll1,
\end{align}
where $E_{\rm uv}$ and $k_{\rm uv}$ are the frequency and momentum scales beyond which the slow collective-field description is not used.

For each permutation $\pi\in\mathfrak S_m$, pair $\P_{a_i}(K_i)$ with $\BP_{b_{\pi(i)}}(K_{\pi(i)}')$ and define
\begin{align}\label{eq:pairing_transfer_def}
    Q_i^{(\pi)}\equiv K_i-K_{\pi(i)}',
    \quad
    \sum_{i=1}^{m}Q_i^{(\pi)}=0,
\end{align}
subject to $\chi_\Lambda\!\left(Q_i^{(\pi)}\right)=1$. These conditions define a slow region $\mathcal S_\pi$ in the momentum space of the vertex. Since $\pi$ can be any permutation of the $m$ outgoing fields, there are $m!$ such pairing regions.

For two distinct permutations $\pi\neq\pi'$, an overlap requires at least one additional momentum-frequency variable to fall within the slow window $\Lambda$. Under the uniform integration measure $\int_K\equiv\int\frac{\dd\omega}{2\pi}\int_{-\pi}^{\pi}\frac{\dd k}{2\pi}$, this extra constraint restricts the phase space, yielding the volume ratio
\begin{align}\label{eq:pairing_sector_overlap}
    \frac{\operatorname{Vol}\left(\mathcal S_\pi\cap\mathcal S_{\pi'}\right)}{\operatorname{Vol}\left(\mathcal S_\pi\right)}=\mathcal O(\lambda),
    \quad
    \pi\neq\pi'.
\end{align}
Thus, the slow pairing sectors are disjoint to leading order in the cutoff expansion.

Within each region, integration over the average variable of a fermion pair produces the slow collective field
\begin{align}\label{eq:appendix_slow_field}
    \hg_{ab}(Q)
    \equiv
    -\i\chi_\Lambda(Q)
    \int_K
    \P_a\!\left(K+\frac{Q}{2}\right)
    \BP_b\!\left(K-\frac{Q}{2}\right).
\end{align}
The contribution from the region $\mathcal S_\pi$ is obtained by replacing every paired product with $\i\hg_{a_i b_{\pi(i)}}$ and including the Grassmann sign $(-1)^{F(\pi)}$ required by that pairing. Consequently, the leading slow part of the vertex has the form
\begin{align}\label{eq:slow_vertex_sector_sum}
    \mathcal V_{2m}^{(<)}[\hg]=\sum_{\pi\in\mathfrak S_m}(-1)^{F(\pi)}\prod_{i=1}^{m}\i\hg_{a_i b_{\pi(i)}}+\mathcal O(\lambda),
\end{align}
where the products include the slow-transfer integrations and the conservation condition in \cref{eq:pairing_transfer_def}.

We define the normalized Gaussian average for an invertible collective matrix field $\hg$ by
\begin{align}\label{eq:gaussian_avg_def}
    \left\langle\mathcal O[\BP,\P]\right\rangle_{\hg}\equiv\frac{1}{\det(-\i\hg^{-1})}\int\D\BP\,\D\P\,e^{\i\BP\hg^{-1}\P}\mathcal O[\BP,\P].
\end{align}
For configurations with zero eigenvalues, this definition is extended through the regulator $\hg\to\hg+\i\delta\I$ with $\delta\to0^+$ or by analytic continuation. The elementary covariance is
\begin{align}\label{eq:gaussian_covariance_G}
    \left\langle\P_a\BP_b\right\rangle_{\hg}=\i\hg_{ab}.
\end{align}
Applying Wick's theorem to \cref{eq:generic_local_2N_vertex} gives
\begin{align}\label{eq:gaussian_all_pairings}
    \left\langle\mathcal V_{2m}[\BP,\P]\right\rangle_{\hg}=\sum_{\pi\in\mathfrak S_m}(-1)^{F(\pi)}\prod_{i=1}^{m}\i\hg_{a_i b_{\pi(i)}}.
\end{align}
Comparison with \cref{eq:slow_vertex_sector_sum} yields
\begin{align}\label{eq:all_channel_equivalence}
    \mathcal V_{2m}^{(<)}[\hg]=\left\langle\mathcal V_{2m}[\BP,\P]\right\rangle_{\hg}+\mathcal O(\lambda).
\end{align}
Therefore, the normalized Gaussian average generates every leading slow pairing region once. The slow-transfer restriction avoids double counting: without it, each permutation would span the full momentum domain, overcounting the vertex by a factor of $m!$.

To verify this correspondence, we expand the bare interaction $\frac{1}{m!}(\BP B \P)^m$ with $B = -2\i\eta\hm_\beta$. At orders $m=2$ and $m=3$, summing the disjoint contractions over the symmetric groups $\mathfrak{S}_2$ and $\mathfrak{S}_3$ with the appropriate Grassmann signs yields
\begin{align}\label{eq:m2_m3_checks}
    \mathcal{V}_4^{(<)}[\hg] &= \frac{1}{2}[\tr(\hg B)]^2 - \frac{1}{2}\tr[(\hg B)^2] + \mathcal{O}(\lambda) \nn= \det(\I+\hg B)\Big|_{m=2} + \mathcal{O}(\lambda), \\
    \mathcal{V}_6^{(<)}[\hg] &= \frac{1}{6}[\tr(\hg B)]^3 - \frac{1}{2}\tr(\hg B)\tr[(\hg B)^2] 
    \nn+ \frac{1}{3}\tr[(\hg B)^3] + \mathcal{O}(\lambda) \nn= \det(\I+\hg B)\Big|_{m=3} + \mathcal{O}(\lambda).
\end{align}
Because $B = -2\i\eta \tau_x^{\rm K}\otimes\mu_\beta^{\rm PH}\otimes\I_R$ is diagonal in replica space, $\tr[(\hg B)^k] = R\,\tr_4[(\hg_0 B_0)^k]$, and the kinematic volume ratio in \cref{eq:pairing_sector_overlap} is independent of $R$. The relative error bound $\mathcal{O}(\lambda)$ is therefore uniform in $R$ and regular as $R\to 1$.

For a local matrix $B$, the Gaussian integral gives $\langle e^{\i\BP B\P}\rangle_{\hg} = \det(\I+\hg B)$. Using $\cosh(2X)=\frac{1}{2}\sum_{\eta=\pm1}e^{2\eta X}$ with $B_{\beta,\eta}=-2\i\eta\hm_\beta$, applying this relation to \cref{eq:all_channel_equivalence} yields
\begin{align}\label{eq:LM_det_appendix}
    \i\L_M^{(<)}[\hg] = 4^{-2R}\sum_{\beta\in\{c,s\}}\sum_{\eta=\pm1}\det_{4R}\left(\I-2\i\eta\hg\hm_\beta\right) - 1 + \mathcal O(\lambda).
\end{align}
The determinant acts on the Keldysh $\otimes$ particle-hole $\otimes$ replica space of dimension $4R$, which satisfies
\begin{align}\label{eq:det_scaling_appendix}
    4^{-2R}\det_{4R}\!\left(\I-2\i\eta\hg\hm_\beta\right)=\det_{4R}\!\left(\frac{\I}{2}-\i\eta\hg\hm_\beta\right).
\end{align}
This relation establishes the absorbed form of the effective Lagrangian density in \cref{eq:LM_det_strict}.

\section{Saddle Point Equation and SCBA}\label{ap:saddle_point}

For a translation-invariant saddle, variation of \cref{eq:S_hs_hg} with respect to $\hs$ and $\hg$, followed by the limit $\epsilon\to0^+$, gives
\begin{subequations}\label{eq:SCBA_appendix}
    \begin{align}
        \hg_{\mathrm{local},0}&=\int_{-\pi}^{\pi}\frac{\dd k}{2\pi}\int_{-\infty}^{\infty}\frac{\dd\omega}{2\pi}\left[G_0^{-1}(k,\omega)+\i\hs_0\right]^{-1},
\label{eq:dyson_saddle}
        \\
        0&=-\i\hs_0+\gamma\left.\frac{\delta(\i\L_M)}{\delta\hg}\right|_{\hg=\hg_{\mathrm{local},0}}.
\label{eq:scba_hs}
    \end{align}
\end{subequations}
The first equation is the symmetric equal-time Green's function defined by \cref{eq:G_reg_def}.

The unconditioned half-filled steady state has a vanishing distribution matrix, so the reference configuration is
\begin{align}
    &\hq_0=\tau_z^{\rm K}\otimes\mu_0^{\rm PH}\otimes\I_R,
    \nn
    \hg_{\mathrm{local},0}=-\frac{\i}{2}\hq_0.
    \label{eq:SCBA_reference}
\end{align}
Varying the determinant in \cref{eq:LM_det_strict} gives
\begin{align}
    \delta(\i\L_M)&=-2\i4^{-2R}\sum_{\beta\in\{c,s\}}
    \sum_{\eta=\pm1}\eta\,\det\!\left( \I_{4R}-2\i\eta\hg\hm_\beta\right) \nn\quad\times\tr\!\left[\hm_\beta\left(\I_{4R}-2\i\eta\hg\hm_\beta\right)^{-1}\delta\hg\right].
    \label{eq:measurement_first_variation}
\end{align}
At the reference configuration, the required matrix relations are
\begin{align}
    &\left(\hq_0\hm_\beta\right)^2=-\I_{4R},
    \nn\left(\I_{4R}-\eta\hq_0\hm_\beta\right)^{-1}=\frac{\I_{4R}+\eta\hq_0\hm_\beta}{2},
    \nn\det\!\left(\I_{4R}-\eta\hq_0\hm_\beta\right)=4^R.
    \label{eq:SCBA_matrix_relations}
\end{align}
Substitution into \cref{eq:measurement_first_variation} yields
\begin{align}
    \left.\delta(\i\L_M)\right|_{\hg_{\mathrm{local},0}}
    =\i4^{1-R}\tr\!\left(\hq_0\delta\hg\right).
    \label{eq:measurement_first_variation_saddle}
\end{align}
The second equation in \cref{eq:SCBA_appendix} then gives
\begin{align}\label{eq:sigma_propto_Q0}
    \hs_0=\gamma4^{1-R}\hq_0\xrightarrow{R\to1}\gamma\hq_0.
\end{align}
The value of the local measurement term at this saddle is
\begin{align}
    \left.\i\L_M \right|_{\hg_{\mathrm{local},0}}=4^{1-R}-1,
\end{align}
which vanishes in the normalization limit $R\to1$.

It remains to verify the first equation in \cref{eq:SCBA_appendix}. Decomposing the Green's function into the two causal sectors gives
\begin{align}
    \hg_{\mathrm{local},0} &=\sum_{q=\pm1}\frac{\tau_0^{\rm K}+q\tau_z^{\rm K}}{2}\otimes\int_{-\pi}^{\pi}\frac{\dd k}{2\pi}\int_{-\infty}^{\infty}\frac{\dd\omega}{2\pi}
    \nn\quad\times\left[\left(\omega+\i q\gamma4^{1-R}\right)\mu_0^{\rm PH}-h_f(k)\right]^{-1}\otimes\I_R.
    \label{eq:SCBA_equal_time_integral}
\end{align}
For every eigenvalue $E$ of $h_f(k)$, the symmetric equal-time prescription gives
\begin{align}
    \int_{-\infty}^{\infty} \frac{\dd\omega}{2\pi}\frac{1}{\omega-E+\i q\gamma4^{1-R}}
    = -\frac{\i q}{2}.
\end{align}
Since the spectral projectors of $h_f(k)$ sum to $\mu_0^{\rm PH}$ and
$\int_{-\pi}^{\pi}\dd k/(2\pi)=1$, \cref{eq:SCBA_equal_time_integral} reduces to
\begin{align}
    \hg_{\mathrm{local},0}
    =-\frac{\i}{2}\tau_z^{\rm K}\otimes \mu_0^{\rm PH}\otimes\I_R
    =-\frac{\i}{2}\hq_0.
\end{align}
This calculation uses the equal-time field generated by the Hubbard--Stratonovich constraint. Taking $\omega\to0$ before the frequency integration would instead evaluate a zero-frequency resolvent. The anomalous $\mu_x^{\rm PH}$ term and its Born-level cancellation therefore do not enter the saddle-point equation.

\section{Measurement-Induced Effective Action}\label{ap:mea}
\subsection{Geometry of Fluctuations on the Parent Manifold}\label{ap:G_hw}
In this subsection, we establish the matrix structure and degree-of-freedom counting of the fluctuation generator $\hw$ on the parent manifold $\mathcal{M}_{\text{parent}}$ and its reduction to $\mathcal{M}_{\text{NLSM}} = \SO(R)$.
In the chiral Keldysh basis, the saddle-point manifold is generated by independent left and right transformations $(u_{\rm L}, u_{\rm R}) \in \mathbb{G}_{\rm CI} = \USp(2R)_{\rm L} \times \USp(2R)_{\rm R}$ acting on the reference matrix $\overline{\hq}_0 = \tau_x^{\rm K} \otimes \mu_0^{\rm PH} \otimes \I_R$.
The stabilizer subgroup is the diagonal group $\mathbb{H}_{\rm CI} = \USp(2R)_{\rm diag}$, yielding the principal chiral orbit $\overline{\hq}[g]$ parameterized by $g = u_{\rm L} u_{\rm R}^{-1} \in \USp(2R)$.
Writing $g = \exp(W_1)$ with Lie algebra generator $W_1 \in \mathfrak{usp}(2R)$, the generator obeys the anti-Hermitian and symplectic conditions
\begin{align}\label{eq:w1_symplectic_app}
    W_1^\dagger = -W_1,
    \quad
    W_1^{\rm T} (\mu_y^{\rm PH}\otimes\I_R) + (\mu_y^{\rm PH}\otimes\I_R) W_1 = 0.
\end{align}
The parent manifold $\mathcal{M}_{\text{parent}} \simeq \USp(2R)$ has real dimension $\dim(\USp(2R)) = R(2R+1)$.

In the chiral basis, the orientational fluctuation takes the form $\overline{\hq}[g] = \exp(\overline{\hw}/2) \overline{\hq}_0 \exp(-\overline{\hw}/2)$ with $\overline{\hw} = \tau_x^{\rm K} \otimes W_1$.
Applying the unitary rotation $V = \frac{1}{\sqrt{2}}(\tau_z^{\rm K} + \tau_x^{\rm K})$ to transform to the causal saddle point $\hq_0 = \tau_z^{\rm K} \otimes \mu_0^{\rm PH} \otimes \I_R$, the fluctuation generator becomes
\begin{align}\label{eq:W_parent_parameterization}
    \hw = V \overline{\hw} V^\dagger = \tau_x^{\rm K} \otimes W_1,
    \quad
    W_1 \in \mathfrak{usp}(2R).
\end{align}
Because Class CI anti-unitary symmetries restrict the chiral orbit to the group field $g \in \USp(2R)$, the generator contains only the $\tau_x^{\rm K}$ component, excluding independent $\tau_y^{\rm K}$ directions.

The low-energy target manifold $\mathcal{M}_{\text{NLSM}} = \SO(R)$ is obtained by freezing massive modes in two steps. First, the charge measurement operator $\hm_c = \tau_x^{\rm K} \otimes \mu_z^{\rm PH} \otimes \I_R$ generates a mass gap for the $\mu_x^{\rm PH}$ and $\mu_y^{\rm PH}$ components via \cref{eq:LM_W_2}, reducing the commutant space to $\mathfrak{u}(R)$ spanned by $\mu_0^{\rm PH}$ and $\mu_z^{\rm PH}$. Second, the pairing gap $\Delta$ gaps out the $\mu_z^{\rm PH}$ channel with mass coefficient $M_\Delta$ [cf.~\cref{ap:pairing_mass}]. The surviving massless Goldstone fluctuations are restricted to
\begin{align}\label{eq:hw_sor_app}
    \hw = \tau_x^{\rm K} \otimes \mu_0^{\rm PH} \otimes \mathcal{R},
    \quad
    \mathcal{R}^{\rm T} = -\mathcal{R} \in \mathfrak{so}(R),
\end{align}
establishing the orthogonal target manifold $\mathcal{M}_{\text{NLSM}} = \SO(R)$ of dimension $R(R-1)/2$.

\subsection{Pairing-Induced Mass for the \texorpdfstring{$\mu_z^{\rm PH}$}{muz} Fluctuation Channel}\label{ap:pairing_mass}
In this subsection, we derive the mass term generated by the superconducting pairing $\Delta \mu_x^{\rm PH}$ for the fluctuation generator $\hw_z = \tau_x^{\rm K} \otimes \mu_z^{\rm PH} \otimes \mathcal{R}_z$. Parameterizing the matrix field as $\hq = e^{\hw/2}\hq_0 e^{-\hw/2} = \hq_0 + \hw\hq_0 + \frac{1}{2}\hw^2\hq_0 + \mathcal{O}(\hw^3)$, the expansion of the trace-log action \cref{eq:S0_hq_final} in the limit $R\to 1$ contains two contributions at second order: 
\begin{align}\label{eq:S0_quad_app} 
\i S_0^{(2)}[\hw] = \frac{\i\gamma}{2}\Tr\left( G \hw^2 \hq_0 \right) + \frac{\gamma^2}{2} \Tr\left( G \hw \hq_0 G \hw \hq_0 \right), 
\end{align} 
where $G = \diag(G^R, G^A)$ is the dressed Green's function at the causal saddle point.

For the Goldstone channel $\hw_0 = \tau_x^{\rm K} \otimes \mu_0^{\rm PH} \otimes \mathcal{R}_0$, evaluating the traces in Keldysh and particle-hole space gives:
\begin{align}
    \i S_0^{(2)}[\hw_0] = &\tr_R(\mathcal{R}_0^2) \int \dd^2\x \int \frac{\dd k \dd\omega}{(2\pi)^2} \nn\times\left[ \frac{\i\gamma}{2}\tr_{\rm PH}(G^R - G^A) - \gamma^2 \tr_{\rm PH}(G^R G^A) \right].
\end{align}
Using the identity $G^R - G^A = -2\i\gamma G^R G^A$, the two terms cancel, ensuring that the $\mu_0^{\rm PH}$ mode remains massless.

For the $\mu_z^{\rm PH}$ fluctuation channel $\hw_z = \tau_x^{\rm K} \otimes \mu_z^{\rm PH} \otimes \mathcal{R}_z$, because $(\mu_z^{\rm PH})^2 = \mu_0^{\rm PH}$, the first term in \cref{eq:S0_quad_app} is unchanged, while the second term contains the particle-hole insertion $\mu_z^{\rm PH}$. The action becomes:
\begin{align}
    \i S_0^{(2)}[\hw_z] = &\gamma^2 \tr_R(\mathcal{R}_z^2) \int \dd^2\x \int \frac{\dd k \dd\omega}{(2\pi)^2} \nn\times\left( \tr_{\rm PH}[G^R G^A] - \tr_{\rm PH}[\mu_z^{\rm PH} G^R \mu_z^{\rm PH} G^A] \right).
\end{align}
Using the Green's functions $G^{R/A}(k,\omega) = [(\omega \pm \i\gamma)\mu_0^{\rm PH} - \xi_k \mu_z^{\rm PH} + \Delta \mu_x^{\rm PH}]^{-1}$ and the cyclic property of the trace, the difference evaluates to: 
\begin{align}\label{eq:trphdeltaomega}
&\tr_{\rm PH}[G^R G^A] - \tr_{\rm PH}[\mu_z^{\rm PH} G^R \mu_z^{\rm PH} G^A]
\en \tr_{\rm PH}\left[ G^R (G^A - \mu_z^{\rm PH} G^A \mu_z^{\rm PH}) \right] \en \frac{4\Delta^2}{\left[ (\omega+\i\gamma)^2 - \xi_k^2 - \Delta^2 \right]\left[ (\omega-\i\gamma)^2 - \xi_k^2 - \Delta^2 \right]}. 
\end{align}
Closing the contour in the upper half-plane, the frequency integral over the poles $\omega = \pm\sqrt{\xi_k^2+\Delta^2} + \i\gamma$ yields:
\begin{align}\label{eq:deltagammaepsilon}
    &\int_{-\infty}^\infty \frac{\dd\omega}{2\pi} \frac{4\Delta^2}{\left[ (\omega+\i\gamma)^2 - \xi_k^2 - \Delta^2 \right]\left[ (\omega-\i\gamma)^2 - \xi_k^2 - \Delta^2 \right]} \en \frac{\Delta^2}{\gamma(\xi_k^2 + \Delta^2 + \gamma^2)}.
\end{align}
Integrating over the energy variable $\xi$ with the normal-state density of states $\nu_F$ gives:
\begin{align}\label{eq:pinudeltagamma}
    \nu_F \int_{-\infty}^\infty \dd\xi \frac{\Delta^2}{\gamma(\xi^2 + \Delta^2 + \gamma^2)} = \frac{\pi \nu_F \Delta^2}{\gamma\sqrt{\Delta^2 + \gamma^2}}.
\end{align}
Using the trace identity $\tr(\hw_z^2) = 4\tr_R(\mathcal{R}_z^2)$, substitution into $\i S_0^{(2)}[\hw_z]$ yields:
\begin{align}\label{eq:S0_quad_pairing_final}
    \i S_0^{(2)}[\hw_z] &= \frac{\pi \nu_F \gamma \Delta^2}{4\sqrt{\Delta^2 + \gamma^2}} \int \dd^2\x\, \tr(\hw_z^2) \nn= \frac{M_\Delta}{2} \int \dd^2\x\, \tr(\hw_z^2),
\end{align}
with the mass coefficient:
\begin{align}\label{eq:M_Delta_app_def}
    M_\Delta = \frac{\pi \nu_F \gamma \Delta^2}{2\sqrt{\Delta^2 + \gamma^2}}.
\end{align}
Because the generator $\hw_z$ is anti-Hermitian ($\tr(\hw_z^2) \le 0$), the Euclidean action $S_E = -\i S_0^{(2)}$ yields a positive energy penalty $S_E = \frac{M_\Delta}{2} \int \dd^2\x\, [-\tr(\hw_z^2)] \ge 0$, confirming that $M_\Delta > 0$ defines a positive mass gap. In the rare-measurement limit $\gamma \ll \Delta$, this evaluates to $M_\Delta \approx \frac{\pi}{2} \nu_F \gamma \Delta$, confirming that the $\mu_z^{\rm PH}$ fluctuations acquire a mass gap and decouple from the low-energy dynamics.

\subsection{Perturbative Expansion of the Measurement Term}\label{ap:mea_f}
We derive the measurement action expanded to the second order in the matrix field $\hq$. We start from the exact determinant representation of the measurement Lagrangian density derived in the main text \cref{eq:LM_det_strict}.
We evaluate the fluctuations around the saddle point $\hq_0$. Instead of a Taylor expansion of the local matrix $\hq \hm_\beta$, which is not small since its eigenvalues are of order unity, we define the expansion in terms of the generator $\hw$. We compute the determinant ratio $\det(\I_{4R} - \eta\hq\hm_\beta)/\det(\I_{4R} - \eta\hq_0\hm_\beta) = \det(\I_{4R} - \hat{W}_{\eta,\beta})$ where $\hat{W}_{\eta,\beta}$ represents the fluctuation matrix:
\begin{align}\label{eq:det_ratio_def}
    \hat{W}_{\eta,\beta} = \eta (\hq - \hq_0) \hm_\beta (\I_{4R} - \eta \hq_0 \hm_\beta)^{-1}.
\end{align}
Using the identity $\det(A) = \exp\left( \Tr \ln A \right)$, the functional determinant becomes:
\begin{align}
    &\det(\I_{4R} - \eta \hq \hm_\beta) \en \det(\I_{4R} - \eta \hq_0 \hm_\beta) \exp\left( \Tr \ln (\I_{4R} - \hat{W}_{\eta,\beta}) \right).
\end{align}
Expanding the logarithm of the determinant to second order in $\hw$ yields the quadratic mass terms. The exponent of the determinant ratio is $E_{\eta,\beta} = \Tr \ln (\I_{4R} - \hat{W}_{\eta,\beta}) \approx -\Tr \hat{W}_{\eta,\beta} - \frac{1}{2}\Tr \hat{W}_{\eta,\beta}^2$.
Expanding the exponential function $\exp(E_{\eta,\beta}) \approx 1 + E_{\eta,\beta} + \frac{1}{2} E_{\eta,\beta}^2$ to second order generates terms of zeroth, first, and second order.

The constant zeroth-order term is determined by the saddle-point determinant $\det_{4R}(\I_{4R} - \eta \hq_0 \hm_\beta) = 4^R$ [cf.~\cref{eq:SCBA_matrix_relations}], yielding the saddle-point action $\i \L_M[\hq_0] = 4^{1-R} - 1$. In the replica limit $R \to 1$, this contribution vanishes, $\i \L_M[\hq_0]\big|_{R\to 1} = 0$, consistent with the Keldysh normalization $Z=1$.

Next, we consider the first-order terms.
The term linear in fluctuations, proportional to $\tr(\hq_0 \hw \hm_\beta)$, vanishes.
This behavior is consistent with $\hq_0$ being a stationary point of the action, which requires the first variation to be zero.
Substituting the Keldysh representation of the saddle point and the fluctuations, we evaluate:
\begin{align}\label{eq:linear_trace_vanish}
    \tr(\hq_0 \hw \hm_\beta) &= \tr \left[ (\tau^{\rm K}_z \otimes \mu^{\rm PH}_0 \otimes \I_R) (\tau^{\rm K}_x \otimes W_1) (\tau^{\rm K}_x \otimes M_\beta) \right] \nn= \tr_K(\tau^{\rm K}_z) \tr(W_1 M_\beta) = 0,
\end{align}
where $\tr_K(\tau^{\rm K}_z) = 0$ ensures the vanishing of the linear contribution, consistent with the saddle-point condition.

For the second-order expansion, using $(\I_{4R} - \eta\hq_0\hm_\beta)^{-1} = \frac{1}{2}(\I_{4R} + \eta\hq_0\hm_\beta)$ and $\delta\hq = \hw\hq_0 + \frac{1}{2}\hw^2\hq_0 + \mathcal{O}(\hw^3)$, the fluctuation matrix $\hat{W}_{\eta,\beta}$ defined in \cref{eq:det_ratio_def} expands as:
\begin{align}
    \hat{W}_{\eta,\beta} = \frac{1}{2}\eta \hw\hq_0\hm_\beta - \frac{1}{2}\hw + \frac{1}{4}\eta \hw^2\hq_0\hm_\beta - \frac{1}{4}\hw^2 + \mathcal{O}(\hw^3).
\end{align}
Expanding the functional determinant to quadratic order via $\ln\det(\I_{4R} - \hat{W}_{\eta,\beta}) = -\Tr \hat{W}_{\eta,\beta} - \frac{1}{2}\Tr \hat{W}_{\eta,\beta}^2$ and summing over $\eta = \pm 1$ eliminates all odd powers of $\eta$. The trace of the squared matrix yields:
\begin{align}
    \sum_{\eta=\pm 1} \Tr(\hat{W}_{\eta,\beta}^2) &= \frac{1}{2} \tr\left[ (\hw\hq_0\hm_\beta)^2 + \hw^2 \right] \nn= \frac{1}{2} \tr\left( \hw\hm_\beta\hw\hm_\beta + \hw^2 \right),
\end{align}
where we used $\hq_0\hm_\beta = -\hm_\beta\hq_0$ and $\hq_0^2 = \I_{4R}$. Combining the contributions gives $\sum_{\eta=\pm 1} [-\Tr \hat{W}_{\eta,\beta} - \frac{1}{2}\Tr \hat{W}_{\eta,\beta}^2] = \frac{1}{4}\tr(\hw^2 - \hw\hm_\beta\hw\hm_\beta) = -\frac{1}{8}\tr([\hw, \hm_\beta]^2)$. Multiplying by the prefactor $4^{-2R} \det_{4R}(\I_{4R} - \eta\hq_0\hm_\beta) = 4^{-R} \xrightarrow{R\to 1} 1/4$ yields the quadratic measurement action:
\begin{align}\label{eq:L_M_2_hw_w}
    \i \L_M^{(2)}[\hw] = -\frac{1}{32} \sum_{\beta\in\{c,s\}} \tr\Bigl( [\hw, \hm_\beta]^2 \Bigr),
\end{align}
which derives the commutator mass term in \cref{eq:LM_W_2}. On the soft subspace $\hw = \tau^{\rm K}_x \otimes W_1$, the spin channel obeys $[\hw, \hm_s] = 0$ identically, leaving the charge channel $\hm_c$ as the sole generator of the mass gap.

\section{Kinetic Terms and Stiffness Coefficients}\label{ap:kin}
\subsection{Spatial Stiffness and Frequency Renormalization}\label{ap:Stiffness}
To extract the stiffness coefficients of the field theory, we perform a gradient expansion of the bare action $S_0[\hq]$ derived in \cref{eq:S0_hq_final}. This requires evaluating the trace-logarithm functional in the presence of the saddle-point self-energy. The slow fluctuations of the matrix field are parameterized through the space-time dependent unitary rotation $\hq(\x) = U(\x) \hq_0 U^\dagger(\x)$. In the following derivation, we retain the explicit coordinate dependence $\x \equiv (t, x)$ for the fluctuation generators $U(\x)$, while the bare propagators are treated in momentum space due to the translational invariance of the saddle point. The renormalized inverse Green's function is expressed as:
\begin{align}\label{eq:renormalized_G}
G^{-1} = G_0^{-1} + \i\hs_0,
\end{align}
which yields the forms for the retarded and advanced sectors:
\begin{align}\label{eq:G_RA}
    G^{R/A}(\omega, k) = \left[ \omega \pm \i\gamma - h_f(k) \right]^{-1}.
\end{align}
Note that while the pairing term $\Delta$ modifies the dispersion relation through $h_f(k)$, the spectral broadening is controlled by the parameter $\gamma$.

To evaluate the spatial kinetic action $S_{\kin}^{(x)}$, we introduce the Hermitian gauge connection $\hat{A}_x = -\i U^\dagger \partial_x U$, which satisfies $\hat{A}_x^\dagger = \hat{A}_x$ and parameterizes the slow spatial gradients of the matrix field. Under minimal coupling on the lattice, the momentum in the hopping term shifts as $k \to k + \hat{A}_x$. Expanding the kinetic component $\xi(k + \hat{A}_x)\mu_z^{\rm PH}$ to second order in $\hat{A}_x$ generates both paramagnetic and diamagnetic vertices:
\begin{align}\label{eq:Vx_definition}
    &\hv_x^{(1)} = (\partial_k \xi_k) K \hat{A}_x = v_k K \hat{A}_x,
    \\
    &\hv_x^{(2)} = \frac{1}{2}(\partial_k^2 \xi_k) K \hat{A}_x^2 = -\frac{1}{2}\xi_k K \hat{A}_x^2,
\end{align}
where $K \equiv \tau_0^{\rm K} \otimes \mu_z^{\rm PH} \otimes \I_R$, $v_k = 2J\sin k$, and $\partial_k^2 \xi_k = -\xi_k = 2J\cos k$. The second-order expansion of the action contains two distinct diagrams:
\begin{align}\label{eq:S2_x_expansion}
    \i \delta S^{(x,2)} &= \i \delta S_{\text{para}}^{(x)} + \i \delta S_{\text{dia}}^{(x)} 
    \nn= -\frac{1}{2}\Tr\left[ G \hv_x^{(1)} G \hv_x^{(1)} \right] - \Tr\left[ G \hv_x^{(2)} \right].
\end{align}
Projecting the fluctuations onto the massless modes satisfying $\{\hat{A}_x, \tau_z^{\rm K}\} = 0$, the paramagnetic bubble diagram evaluates to:
\begin{align}\label{eq:S2_simplified}
    \i \delta S_{\text{para}}^{(x)} =& -\frac{1}{2}\int \dd^2\x\, \tr_{\text{K,R}}(\hat{A}_x^2) \nn\times\int \frac{\dd k \dd\omega}{(2\pi)^2} v_k^2 \tr_{\text{PH}}\left[ \mu_z^{\rm PH} G^R(\omega, k) \mu_z^{\rm PH} G^A(\omega, k) \right].
\end{align}
Evaluating the frequency integral over the particle-hole trace yields:
\begin{align}\label{eq:trace_nambu_calc}
    &\int_{-\infty}^\infty \frac{\dd\omega}{2\pi}\tr_{\text{PH}}\left[ \mu_z^{\rm PH} G^R(\omega, k) \mu_z^{\rm PH} G^A(\omega, k) \right]
    \nn= \frac{\xi_k^2 + \gamma^2}{\gamma(\xi_k^2 + \Delta^2 + \gamma^2)} = \frac{1}{\gamma} - \frac{\Delta^2}{\gamma(\xi_k^2 + \Delta^2 + \gamma^2)}.
\end{align}
The first term gives an ultraviolet contribution $\mathcal{I}_{\text{para}}^{\text{UV}} = \frac{1}{\gamma}\int_{-\pi}^\pi \frac{\dd k}{2\pi} v_k^2 = \frac{2J^2}{\gamma}$. Concurrently, the diamagnetic contact diagram evaluates to:
\begin{align}\label{eq:dia_eval}
    \i \delta S_{\text{dia}}^{(x)} &= \frac{1}{2}\int \dd^2\x\, \tr_{\text{K,R}}(\hat{A}_x^2) \int_{-\pi}^\pi \frac{\dd k}{2\pi}\frac{\xi_k^2}{\gamma} 
    \nn= \frac{1}{2}\int \dd^2\x\, \tr_{\text{K,R}}(\hat{A}_x^2) \left(\frac{2J^2}{\gamma}\right).
\end{align}
By the Ward identity, the ultraviolet constant $\frac{2J^2}{\gamma}$ cancels between the paramagnetic and diamagnetic terms. The remaining low-energy spatial kernel is:
\begin{align}\label{eq:integral_evaluation}
    \mathcal{I}_x &= \int_{-\pi}^\pi \frac{\dd k}{2\pi} \frac{v_k^2 \Delta^2}{\gamma(\xi_k^2 + \Delta^2 + \gamma^2)}
    \nn\approx \nu_F \int_{-\infty}^\infty \dd\xi \frac{v_F^2 \Delta^2}{\gamma(\xi^2 + \Delta^2 + \gamma^2)}
    \nn= \frac{\pi \nu_F v_F^2 \Delta^2}{\gamma \sqrt{\Delta^2 + \gamma^2}},
\end{align}
where we linearized the spectrum near the Fermi points in the regime $\Delta, \gamma \ll J$, with $\nu_F = 1/(2\pi J)$ and $v_F = 2J$. Using $\tr_{\text{K,R}}(\hat{A}_x^2) = \frac{1}{8}\tr((\partial_x \hq)^2)$, the spatial kinetic action takes the form:
\begin{align}\label{eq:final_spatial_action}
    \i S_{\kin}^{(x,2)}[\hq] &= -\frac{1}{16}\left[ \frac{\pi \nu_F v_F^2 \Delta^2}{\gamma \sqrt{\Delta^2 + \gamma^2}} \right] \int \dd^2\x\, \tr\left[ (\partial_x \hq)^2 \right]
    \nn= -\frac{\pi \nu_F D_x}{8} \int \dd^2\x\, \tr\left[ (\partial_x \hq)^2 \right].
\end{align}
Matching the coefficients identifies the spatial stiffness coefficient:
\begin{align}\label{eq:D_final_def}
    D_x = \frac{v_F^2 \Delta^2}{2\gamma \sqrt{\Delta^2 + \gamma^2}} \approx \frac{v_F^2 \Delta}{2\gamma} \quad (\gamma \ll \Delta).
\end{align}
The spatial stiffness is governed by the subgap states generated by measurement broadening within the pairing gap.

The frequency-dependent component of the action originates from the variation of the fermion measure in the path integral formulation. This contribution is identified with the first-order vacuum polarization diagram:
\begin{align}\label{eq:S_freq_trace}
    \i \delta S^{(t,1)} = -\i\Tr \left( G \hv_t \right),
\end{align}
where $\hv_t = -U^\dagger \partial_t U$ denotes the temporal gauge potential derived from the unitary rotation.
Evaluating the trace requires including the matrix structure of the BCS Green's function. Concretely, we require the difference between the retarded and advanced traces to couple to the matrix field $\hq_0$. Using the relations $\tr_{\text{PH}}(G^R) = \frac{-2\i\gamma}{\xi^2 + \Delta^2 +\gamma^2}$ and $\tr_{\text{PH}}(G^A) = \frac{+2\i\gamma}{\xi^2 + \Delta^2 +\gamma^2}$, the relevant integral kernel $\mathcal{I}_{t1}$ is calculated as:
\begin{align}\label{eq:I_t_1}
\mathcal{I}_{t1}&= \nu_F \int_{-\infty}^{\infty} \dd\xi \left[\tr_{\text{PH}}(G^R) -\tr_{\text{PH}}(G^A) \right]
\nn= -4\i\gamma \nu_F \int_{-\infty}^{\infty} \frac{\dd\xi}{\xi^2 + \Delta^2 +\gamma^2} = \frac{-4\i\pi\gamma\nu_F}{\sqrt{\Delta^2 + \gamma^2}}.
\end{align}
Substituting this kernel back into the expansion functional yields the temporal kinetic term:
\begin{align} \label{eq:final_temporal_action} 
    \i S_{\kin}^{(t,1)}[\hq] = \frac{\pi \nu_F Z_\omega}{2} \int \dd^2\x\,  \tr \left( \hq_0 U^\dagger \partial_t U \right). 
\end{align}
Here, the dynamics are governed by the frequency renormalization factor:
\begin{align}\label{eq:Z_omega}
    Z_\omega \equiv \cfrac{2\gamma}{\sqrt{\Delta^2+\gamma^2}}.
\end{align}
The coefficient of this term reflects the residual density of states at the Fermi level induced by the measurement broadening within the BCS gap. In the limit $\Delta \to 0$, the expression recovers the standard metallic result, whereas for $\Delta \gg \gamma$, the dynamics are frozen out. The suppression of the diffusive $z=2$ contribution is a consequence of the algebraic cancellation within the $\SO(R)$ manifold. The trace structure ensures that the linear frequency term vanishes, identifying the second-order derivative as the leading temporal term that recovers $z=1$ scaling.

Combining the spatial and temporal components derived in the preceding steps, and using the identity $ \tr(\hq_0 U^\dagger \partial_t U) = \frac{1}{2} \tr(\hq_0 \hq \partial_t \hq)$ to unify the representation, the full kinetic action is established as:
\begin{align}\label{eq:S_kin}
\i S_{\kin}[\hq] = -\frac{\pi \nu_F}{8} \int \dd^2\x\,  \tr \left[ D_x (\partial_x \hq)^2 - 2 Z_\omega \hq_0 \hq \partial_t \hq \right].
\end{align}
While the unprojected action on the parent manifold contains the usual first-order time derivative term representing diffusive dynamics, after projecting onto the $\SO(R)$ manifold this term vanishes, and the leading dynamics is governed by the second-order temporal term, yielding a $z=1$ scaling.
However, the dynamical scaling ($z=1$ vs. $z=2$) depends on the geometric properties of the NLSM manifold, which determines whether the first-order time derivative term survives. We will address the specific action on the identified manifold in \cref{sec:WZW}.

\subsection{Vanishing of the Berry Phase Term}\label{ap:vanish}
We evaluate the linear temporal kinetic term $\i S_{\kin}^{(t,1)}[\hq]$ [cf.~\cref{eq:final_temporal_action}] on the massless manifold $\mathcal{M}_{\text{NLSM}} = \SO(R)$. Parameterizing the matrix field as $\hq = U \hq_0 U^\dagger$ with $U = \exp(\hw/2)$, the fluctuation generator $\hw$ assumes the form:
\begin{align} \label{eq:W_SOR_structure}
\hw(\x, t) = \tau^{\rm K}_x \otimes \mu^{\rm PH}_0 \otimes \mathcal{R}(\x, t).
\end{align}
Here, the Pauli matrix $\tau^{\rm K}_x$ acts on the Keldysh space, while $\mu^{\rm PH}_0$ denotes the identity matrix in the particle-hole space, which commutes with the pairing and measurement terms. The matrix $\mathcal{R}$ is real and antisymmetric, generating the $\SO(R)$ rotations in the replica space:
\begin{align} \label{eq:R_antisymmetry}
\mathcal{R}^{\rm T} = -\mathcal{R}, \quad \mathcal{R} \in\mathfrak{so}(R).
\end{align}
The saddle point is given by $\hq_0 = \tau^{\rm K}_z \otimes \mu^{\rm PH}_0 \otimes \I_R$. Expanding $U^\dagger \partial_t U$ in powers of the fluctuation generator $\hw$ yields:
\begin{align}\label{eq:UtU}
 U^\dagger \partial_t U &= e^{-\hw/2} \partial_t e^{\hw/2} \nn= \frac{1}{2} \partial_t \hw - \frac{1}{8} [\hw, \partial_t \hw] + \mathcal{O}(\hw^3).
\end{align}
Substituting the expansion of $U^\dagger \partial_t U$ into the first-order temporal action \cref{eq:final_temporal_action} produces products of the generator $\hw$ and its time derivatives. On the projected manifold, the fluctuation generator satisfies $\hw \propto \tau^{\rm K}_x \otimes \mu^{\rm PH}_0 \otimes \mathcal{R}$ [cf.~\cref{eq:W_SOR_structure}]. Consequently, a product of $n$ generators carries the Keldysh factor $(\tau^{\rm K}_x)^n$, which equals $\tau^{\rm K}_x$ for odd $n$ and $\tau^{\rm K}_0$ for even $n$. Coupling to the causal saddle point $\hq_0 \propto \tau^{\rm K}_z$ involves evaluating the Keldysh trace:
\begin{align}\label{eq:T1tt}
\tr_K[\tau^{\rm K}_z (\tau^{\rm K}_x)^n] = 
\begin{cases} 
\tr_K(\tau^{\rm K}_z \tau^{\rm K}_x) = 0, & n \text{ odd}, \\ 
\tr_K(\tau^{\rm K}_z) = 0, & n \text{ even}. 
\end{cases}
\end{align}
Because this trace vanishes for every positive integer $n$, the first-order temporal kinetic action vanishes to all orders on the $\SO(R)$ manifold:
\begin{align} \label{eq:temporal_vanish_proof}
\i S_{\kin}^{(t,1)}[\hq]\Big|_{\hq \in \SO(R)} \equiv 0.
\end{align}
This cancellation suppresses the conventional diffusive $z=2$ mode. The leading temporal dynamics are instead governed by the second-order expansion, yielding a linear dispersion $\omega \propto k$ with dynamical critical exponent $z=1$ and effective spacetime dimension $D_{\text{eff}} =2$, where the coupling constant is marginal.

\subsection{Compressibility and Second-Order Dynamics}\label{ap:second_t}
Given the vanishing of the first-order temporal term, the leading dynamical contribution arises from the second-order expansion of the kinetic action.
This term corresponds to the vacuum polarization bubble and determines the temporal stiffness, or generalized compressibility, of the effective field theory.
The coefficient for the $(\partial_t \hq)^2$ term is calculated as follows.

We start from the expansion of the Keldysh action $\i S = \Tr \ln G^{-1}$.
Under the local unitary rotation $\hq = U \hq_0 U^\dagger$, the inverse Green's function transforms as $G^{-1} \to G^{-1} - \hat{A}_t$, where $\hat{A}_t = U^\dagger \partial_t U$ is the anti-Hermitian temporal gauge connection satisfying $\hat{A}_t^\dagger = -\hat{A}_t$. Expanding the action to second order in $\hat{A}_t$, we obtain: 
\begin{align}\label{eq:S_kin_define} 
\i S_{\kin}^{(t,2)} = -\frac{1}{2} \Tr \left( G \hat{A}_t G \hat{A}_t \right). 
\end{align}
In the slow-field limit where the characteristic frequency of $\hat{A}_t(\x)$ is much smaller than the relaxation rate $\gamma$, the action simplifies to the static response:
\begin{align}\label{eq:tr_K_GAGA}
    \i S_{\kin}^{(t,2)} = -\frac{1}{2} \int \dd^2\x\, \tr_{\text{K,R}}(\hat{A}_t^2) \cdot \mathcal{I}_{t2},
\end{align}
where the static polarization kernel $\mathcal{I}_{t2}$ is defined by:
\begin{align}\label{eq:I_t_2_define}
    \mathcal{I}_{t2} = \int \frac{\dd k \dd\omega}{(2\pi)^2} \tr_{\text{PH}} \left[ G^R(\omega, k) G^A(\omega, k) \right].
\end{align}
The Green's functions in the particle-hole basis read:
\begin{align}\label{eq:G_A_R}
    G^{R/A}(\omega, k) = [(\omega \pm \i\gamma) \mu^{\rm PH}_0 - \xi_k \mu^{\rm PH}_z + \Delta \mu^{\rm PH}_x]^{-1}.
\end{align}
The product of the Green's functions evaluates to:
\begin{align}
    G^R(\omega, k) G^A(\omega, k) =& \frac{(\omega+\i\gamma)\mu_0^{\rm PH} + \xi_k \mu_z^{\rm PH} - \Delta \mu_x^{\rm PH}}{(\omega+\i\gamma)^2 - \xi_k^2 - \Delta^2} \nn\cdot\frac{(\omega-\i\gamma)\mu_0^{\rm PH} + \xi_k \mu_z^{\rm PH} - \Delta \mu_x^{\rm PH}}{(\omega-\i\gamma)^2 - \xi_k^2 - \Delta^2}.
\end{align}
Evaluating the trace over the particle-hole space and performing the contour integration over $\omega$ yields $\int \frac{\dd\omega}{2\pi}\tr_{\rm PH}[G^R(\omega, k) G^A(\omega, k)] = 1/\gamma$. Integrating this result over the one-dimensional Brillouin zone gives the polarization kernel:
\begin{align}\label{eq:I_t2_final}
    \mathcal{I}_{t2} = \int_{-\pi}^\pi \frac{\dd k}{2\pi}\left(\frac{1}{\gamma}\right) = \frac{1}{\gamma}.
\end{align}
Using the relation $\partial_t \hq = 2U \hat{A}_t \hq_0 U^\dagger$, the identity $(\partial_t \hq)^2 = -4U \hat{A}_t^2 U^\dagger$ gives $\tr(\hat{A}_t^2) = -\frac{1}{4}\tr((\partial_t \hq)^2)$. Tracing over the Keldysh and replica spaces yields $\tr_{\text{K,R}}(\hat{A}_t^2) = \frac{1}{2}\tr(\hat{A}_t^2) = -\frac{1}{8}\tr((\partial_t \hq)^2)$. Substituting $\mathcal{I}_{t2}$ and $\tr_{\text{K,R}}(\hat{A}_t^2)$ into \cref{eq:tr_K_GAGA} leads to:
\begin{align}\label{eq:iS_kin_t_2}
\i S_{\kin}^{(t,2)} &= \frac{1}{16\gamma} \int \dd^2\x\, \tr\left[ (\partial_t \hq)^2 \right] \nn= \frac{\pi \nu_F D_t}{8} \int \dd^2\x\, \tr\left[ (\partial_t \hq)^2 \right],
\end{align}
which determines the temporal stiffness coefficient:
\begin{align}\label{eq:temporal_action_final}
D_t = \frac{1}{2\pi \nu_F \gamma} = \frac{J}{\gamma}.
\end{align}
This result establishes the coefficient $D_t$ in \cref{eq:S_kin_final}.

\section{Momentum Shell Decomposition and Effective Action}\label{ap:momentumshell}
The coefficients of the NLSM are derived from the Hamiltonian parameters of the monitored BCS chain. For the tight-binding dispersion $h(k) = -2J \cos k$ and pairing amplitude $\Delta$, the quasiparticle energy is $E_k = \sqrt{h(k)^2 + \Delta^2}$. The low-energy dynamics are governed by the excitations near the Fermi points, where the Fermi velocity $v_F$ is defined:
  \begin{align}
      v_F = \left| \partial_k \xi_k \right|_{k_F} = 2J.
  \end{align}
In the presence of a finite measurement rate $\gamma$, the quasiparticle lifetime is limited to $\tau = (2\gamma)^{-1}$. The resulting mean free path $l_0$ gives the spatial scale for the diffusive fluctuations:
\begin{align}\label{eq:mean_free_path_expression}
      l_0 = v_F \tau = \frac{v_F}{2\gamma}.
\end{align}
This length scale defines the ultraviolet cutoff $\Lambda = l_0^{-1} = \frac{2\gamma}{v_F}$ for the momentum-shell integrations.
To isolate the slow rotational modes from the invariant matrix structures, we map the gradient action of $\hq$ to the $\SO(R)$ group manifold. Substituting the generator $\hw = \tau^{\rm K}_x \otimes \mu^{\rm PH}_0 \otimes \mathcal{R}$ into the parameterization $\hq = \exp(\hw/2) \hq_0 \exp(-\hw/2)$ yields the Taylor expansion $\exp(\hw/2) = \cosh(\mathcal{R}/2) + \tau^{\rm K}_x \sinh(\mathcal{R}/2)$, using $(\tau_x^{\rm K})^2 = \I_2$. Because $\mathcal{R}$ is a real antisymmetric matrix ($\mathcal{R}^{\rm T} = -\mathcal{R}$), its non-zero eigenvalues are purely imaginary conjugate pairs $\pm \i \theta_j$, so the matrix functions on each invariant block reduce to trigonometric functions. Anti-commuting $\tau_x^{\rm K}$ with the saddle point $\hq_0 \propto \tau_z^{\rm K}$ simplifies the matrix field to $\hq = \tau_z^{\rm K}\cosh(\mathcal{R}) - \i \tau_y^{\rm K}\sinh(\mathcal{R})$, which corresponds to $\tau_z^{\rm K}\cos\theta_j + \tau_y^{\rm K}\sin\theta_j$ on each rotation block. Computing the squared gradient $\tr_{4R}(\partial_\mu \hq)^2$ traces over the constant Keldysh and particle-hole subspaces, yielding $\tr_{4R}[(\partial_\mu \hq)^2] = 4 \tr[(\partial_\mu Y)(\partial_\mu Y^{\rm T})]$ with the rotational field $Y = \exp(\mathcal{R}) \in \SO(R)$. Exploiting the orthogonality condition $Y^{-1} = Y^{\rm T}$, we express this energy density in terms of the squared current:
\begin{align}\label{eq:kinetic_to_current}
\tr \left[ (\partial_\mu Y)(\partial_\mu Y^{\rm T}) \right] = - \tr ( J_\mu^2 ).
\end{align}
Here, the current $J_\mu = Y^{\rm T}\partial_\mu Y$ is a real antisymmetric matrix spanning the Lie algebra $\mathfrak{so}(R)$, which satisfies $\tr(J_\mu^2) \le 0$. Consequently, the kinetic density $-\tr(J_\mu^2) \ge 0$ is positive semi-definite. Under the parametrization $Y = Y_0 e^{\mw}$, the total current relates to the background connection $A_\mu = Y_0^{-1}\partial_\mu Y_0$ through the gauge transformation:
\begin{align}
J_\mu = e^{-\mw} A_\mu e^\mw + e^{-\mw} \partial_\mu e^\mw.
\end{align}
Applying the Baker-Campbell-Hausdorff formula and expanding to the second order in $\mw$, we obtain:
\begin{align} \label{eq:current_expansion}
J_\mu &= A_\mu + \partial_\mu \mw + [A_\mu, \mw] \nn
\quad - \frac{1}{2} [\mw, \partial_\mu \mw] + \frac{1}{2} [[A_\mu, \mw], \mw] + \mathcal{O}(\mw^3).
\end{align}
To streamline the notation, we introduce the covariant derivative with respect to the background field, $\D_\mu \mw = \partial_\mu \mw + [A_\mu, \mw]$. Thus, the expansion assumes the compact form:
\begin{align}
J_\mu \approx A_\mu + \D_\mu \mw - \frac{1}{2}[\mw, \D_\mu \mw].
\end{align}
We substitute the current expansion into the kinetic density $-\tr(J_\mu^2)$ to isolate the terms quadratic in $\mw$. The term linear in fluctuations, $\tr(A_\mu \D_\mu \mw)$, vanishes by the background equation of motion $\D_\mu A_\mu = 0$. Expanding $(\D_\mu \mw)^2 = (\partial_\mu \mw + [A_\mu, \mw])^2$ together with the cross-term $-\tr(A_\mu [\mw, \D_\mu \mw])$ leads to the cancellation of the curvature-induced mass terms $\tr([A_\mu, \mw]^2)$, yielding
\begin{align}\label{eq:curvature_cancel}
-\tr(J_\mu^2)\big|_{\mathcal{O}(\mw^2)} = -\tr(\partial_\mu \mw)^2 - \tr([A_\mu, \mw] \partial_\mu \mw).
\end{align}
This algebraic cancellation ensures that the soft modes remain massless. We define the normalization of the generator $T^a$ by $\tr(T^a T^b) = -2\delta^{ab}$. Under this convention, the quadratic term in the action has a positive coefficient, ensuring a well-defined Euclidean path integral with proper weight $\exp(-S_{\eff})$.

Incorporating the coupling constant defined in the main text, we obtain the resulting second-order fluctuation action:
\begin{align}\label{eq:Action_Expanded_Final}
S^{(2)}[\mw] = \frac{1}{g} \int \dd^2 \bm{r} \,\tr \left[ -(\partial_\mu \mw)^2 - [A_\mu, \mw] \partial_\mu \mw \right].
\end{align}
Given the antisymmetric nature of $\mw \in \mathfrak{so}(R)$, the kinetic term $-\tr(\partial_\mu \mw)^2$ is positive definite, ensuring stability. 
With the fluctuation action established in \cref{eq:Action_Expanded_Final}, we identify the interaction vertex as
\begin{align}
    S_{\text{int}} = -\frac{1}{g} \int \dd^2 \bm{r} \, \tr \left( [A_\mu, \mw] \partial_\mu \mw \right).
\end{align}
The one-loop correction is governed by the second cumulant expansion $\Delta S^{\text{RG}} \sim -\langle S_{\text{int}}^2 \rangle_0$.
This contribution involves the contraction of two interaction vertices:
\begin{align}\label{eq:S_RG_define_integral}
    \Delta S^{\text{RG}} &\sim -  \frac{1}{g^2} \int \dd^2 \bm{r} \dd^2 \bm{r}'  \left\langle \mathcal{F}(\bm{r}) \mathcal{F}(\bm{r}^\prime) \right\rangle_0, \n
    \mathcal{F}(\bm{r}) &\equiv \tr \left( [A_\mu(\bm{r}), \mw(\bm{r})] \partial_\mu \mw(\bm{r}) \right).
\end{align}
To perform the momentum-shell integration, the fast fluctuations $\mw$ are assigned loop momenta $\q$ within the shell $\Lambda/b \le |\q| < \Lambda$, while the background connection $A_\mu$ varies on longer length scales and is treated as constant over the fast-mode integration. Expanding the fluctuation field in terms of the $\mathfrak{so}(R)$ generators, $\mw = \sum_a \mw_a T^a$ with $a \in \{1, \dots, R(R-1)/2\}$, and adopting the normalization $\tr(T^a T^b) = -2\delta^{ab}$, the Gaussian action evaluates to $S_0[\mw] = \int \frac{\dd^2\q}{(2\pi)^2} \frac{2q^2}{g} \sum_a |\mw_a(\q)|^2$. The corresponding propagator for the scalar components reads:
\begin{align}\label{eq:propagator_def}
\langle \mw_a(\q) \mw_b(-\q) \rangle_0 \sim \frac{g}{q^2} \delta_{ab}.
\end{align}
The expectation value $\langle S_{\text{int}}^2 \rangle_0$ entails the contraction of four Lie algebra generators. Parity of the momentum integral causes the single-derivative expectation value $\langle \mw_a \partial_\mu \mw_b \rangle_0$ to vanish. The logarithmic contribution arises from the contraction of two derivative terms:
\begin{align}\label{eq:derivative_contraction}
    \langle \partial_\mu \mw_a(\q) \partial_\nu \mw_b(-\q) \rangle_0 = \frac{g q_\mu q_\nu}{4q^2} \delta_{ab}.
\end{align}
The commutator expands as $[T^a, \mw] = \sum_c \mw_c [T^a, T^c] = \sum_{c,d} f^{acd} \mw_c T^d$, where $f^{acd}$ are the structure constants of $\mathfrak{so}(R)$ satisfying $\tr([T^a, T^c] T^d) = -2 f^{acd}$. Evaluating the Wick contractions over the two equivalent channels and summing the structure constants via $\sum_{c,d} f^{acd} f^{bcd} = 2(R-2) \delta^{ab}$ yields 
\begin{align}\label{eq:tgWW}
\langle \tr( [T^a, \mw] \partial_\mu \mw ) \tr( [T^b, \mw] \partial_\nu \mw ) \rangle_0 = g^2(R-2) \delta^{ab} \frac{q_\mu q_\nu}{q^4}. 
\end{align}
The momentum-shell integration over the fast modes gives
\begin{align}
    \int_{\Lambda/b}^{\Lambda} \frac{\dd^2 \q}{(2\pi)^2} \frac{q_\mu q_\nu}{q^4} \sim \delta_{\mu\nu} \ln b.
\end{align}
Evaluating the angular integral maps the derivative product to its isotropic average, while the radial momentum integration generates the logarithmic factor $\ln b$, finally gives the one-loop quantum correction: 
\begin{align}\label{eq:Delta_RG_cbeta}
\Delta S^{\text{RG}} = - \left[ c_\beta(R-2) \ln b \right] \int \dd^2 \bm{r} \,\tr [(\nabla Y_0)(\nabla Y_0^{\rm T})], \end{align} 
where the one-loop momentum-shell coefficient is identified as $c_\beta = \frac{1}{16\pi}$. We determine the renormalized coupling constant $g(b)$ by adding this fluctuation correction to the bare background action, which gives the running coupling relation $1/g(b) = 1/g - c_\beta(R-2)\ln b$.

\section{Entanglement Entropy from Twist Operators}\label{ap:EE}

We evaluate the entanglement entropy of a spatial subsystem $A = [0, L]$ by inserting twist operators at its boundary endpoints. In polar coordinates $(r, \theta)$ centered on a boundary defect, a $2\pi$ rotation imposes the monodromy condition on the $\SO(R)$ matrix field:
\begin{align}\label{eq:Y_vortex_ansatz}
    Y(r, \theta + 2\pi) = \Omega_N Y(r, \theta).
\end{align}
Here, $\Omega_N$ denotes the cyclic permutation matrix acting on the $N$ replica copies. For odd integers $N = 3, 5, 7, \dots$, the permutation matrix satisfies $\det \Omega_N = +1$ and lies within the connected component $\SO(N) \subset \SO(R)$. The classical saddle-point configuration that minimizes the Euclidean action under this boundary condition is given by the vortex ansatz:
\begin{align}\label{eq:vortex_ansatz}
    Y_{\rm cl}(r, \theta) = e^{\theta \hat{X}} Y_0,
\end{align}
where $Y_0$ is a constant reference matrix and $\hat{X} \in \mathfrak{so}(N)$ is the Lie algebra generator satisfying $\exp(2\pi \hat{X}) = \Omega_N$.

In polar coordinates, the spatial gradient of the vortex configuration evaluates to $\nabla Y_{\rm cl} = \frac{1}{r} \hat{X} Y_{\rm cl} \hat{\bm{e}}_\theta$. The kinetic energy density is obtained by evaluating the trace:
\begin{align}\label{eq:tr_Y_cl_Y_cl}
   \tr [ (\nabla Y_{\rm cl}) (\nabla Y_{\rm cl}^{\rm T}) ] = -\frac{1}{r^2} \tr(\hat{X}^2).
\end{align}
Because $\hat{X}$ is a real antisymmetric matrix, the eigenvalues of $\hat{X}$ are purely imaginary, ensuring $-\tr(\hat{X}^2) \ge 0$.

For subsystem $A=[0,L]$, the twist defects at the two endpoints form a vortex dipole whose topological charges cancel for $r \gg L$, screening the far-field gradients. Truncating the single-vortex logarithmic integral at the separation scale $L$ and incorporating the running coupling $g(\ell)$ with $\ell(r) = \ln(r/l_0)$, the free energy of the two boundary defects evaluates to:
\begin{align}\label{eq:Delta_F_N_int}
 \Delta F_N(L) \sim - \tr(\hat{X}^2) \int_{0}^{\ell_L} \frac{\dd \ell}{g(\ell)}.
\end{align}
The cyclic permutation generator $\hat{X} \sim \ln \Omega_N \in \mathfrak{so}(N)$ provides a positive vortex energy contribution $-\tr(\hat{X}^2) \sim (N-1) > 0$. Evaluating the replica derivative $\se(L) = \lim_{N \to 1} \frac{\partial \Delta F_N(L)}{\partial N}$ with the running stiffness $\frac{1}{g(\ell)} = \frac{1}{g_0} + c_\beta \ell$ yields the two-component scaling form:
\begin{align}\label{eq:entropy_final_app} 
\se(L) \approx A \left[ \ln\left(\frac{L}{l_0}\right) \right]^2 + B \ln\left(\frac{L}{l_0}\right) + \mathcal{O}(1),
\end{align} 
where $A \sim c_\beta \sim \mathcal{O}(1) > 0$ is a universal coefficient governed by the one-loop flow parameter, and $B \sim 1/g_0$ denotes the non-universal slope determined by the effective bare stiffness, establishing \cref{eq:entropy_microscopic_expanded} in the main text.

The derivation of \cref{eq:entropy_final_app} from the lattice Hamiltonian relies on a hierarchy of approximations. In the continuum field theory, the lattice dispersion is linearized near the Fermi points, the Poissonian measurements are coarse-grained into the self-energy $\hs_0 = \gamma \hq_0$ within the self-consistent Born approximation, and spacetime anisotropy is absorbed into the coordinate scaling $x_0 = \sqrt{D_x/D_t}\tau$. Furthermore, fluctuation channels with pairing mass $M_\Delta$ are frozen at tree level, neglecting finite correlation length corrections $\xi_{\text{mass}} \sim \sqrt{D_x/M_\Delta}$. In the infrared limit, the beta function is evaluated at one-loop order, while the twist defect configuration is evaluated using non-interacting classical vortices that omit microscopic core structures and multipole interactions between the boundaries.

Due to these approximations, the pure $\ln^2 L$ scaling characterizes the asymptotic infrared regime where $\ln(L/l_0) \gg B/A$. The crossover length scale $L_{\text{cross}}$ above which the $\ln^2 L$ term exceeds the linear $\ln L$ term is parameterized as:
\begin{align}\label{eq:L_cross_estimate}
L_{\text{cross}} \sim l_0 \exp\left( \frac{B}{A} \right) \sim l_0 \exp\left( \frac{\mathcal{O}(1)}{g_0} \right).
\end{align}
Because the ratio $B/A$ depends on the non-universal bare coupling $g_0$, \cref{eq:L_cross_estimate} provides a parametric crossover scale rather than a threshold. For typical parameters in numerical simulations, $\ln(L_{\text{cross}}/l_0)$ exceeds the scale window accessible on finite lattices. Consequently, numerical data probe the crossover regime, where the effective slope $\lambda_{\gamma,\Delta}$ extracted from finite-size fitting exhibits parametric dependence on $(\Delta, \gamma)$ through the linear logarithmic term.

\end{document}